\documentclass[sigconf, screen, acmsmall, table]{acmart} 
\usepackage{filecontents}
\usepackage[utf8]{inputenc}
\usepackage{multirow}
\usepackage{array}
\usepackage{caption}
\usepackage{geometry}
\usepackage[normalem]{ulem}
\usepackage{booktabs}
\usepackage{makecell}
\usepackage[table]{xcolor}
\usepackage{graphicx}
\usepackage{ragged2e}
\usepackage{colortbl}
\usepackage{lscape}
\usepackage{subcaption}
\newcommand{\bc}[1]{\textcolor{black}{#1}}
\newcommand{\oc}[1]{\textcolor{black}{#1}}
\usepackage[dvipsnames, xcdraw]{xcolor}
\usepackage{tcolorbox}

\definecolor{LightBlue}{HTML}{A8D8FF}      
\definecolor{LightGreen}{HTML}{B6F2C1}     
\definecolor{LightYellow}{HTML}{FFF5A3}    
\definecolor{LightPink}{HTML}{FFB7C5}      
\definecolor{LightLavender}{HTML}{D1C4FF}  
\definecolor{LightPeach}{HTML}{FFCBA4}     

\AtBeginDocument{%
  \providecommand\BibTeX{{%
    \normalfont B\kern-0.5em{\scshape i\kern-0.25em b}\kern-0.8em\TeX}}}

\setcopyright{acmlicensed}
\copyrightyear{2026}
\acmYear{2026}
\acmDOI{XXXXXXX.XXXXXXX}

\acmConference[Accepted at CSCW 2026]{The 29th ACM SIGCHI Conference on Computer-Supported Cooperative Work \& Social Computing}{October 10--14, 2026}{Salt Lake City, UT, USA}
\acmISBN{978-1-4503-XXXX-X/18/06}

\begin{document}

\title[Exploring the Perception Gaps between Interviewers and Interviewees of the AI-Driven Asynchronous Interviewers]{Experience and Adaptation in AI Recruitment Systems: A Combined Analysis of Online Discourse and Interface Design}

\title[]{Expecting Too Much, Getting Too Little: Exploring the Challenges and Design Opportunities of Asynchronous AI Interviewers}


\author{Md Nazmus Sakib}
\email{msakib1@umbc.edu}
\orcid{0009-0003-8282-3931}
\affiliation{%
  \institution{University of Maryland, Baltimore County}
  \city{Baltimore}
  \state{Maryland}
  \country{USA}
}
\author{Naga Manogna Rayasam}
\email{tp34657@umbc.edu}
\orcid{}
\affiliation{%
  \institution{University of Maryland, Baltimore County}
  \city{Baltimore}
  \state{Maryland}
  \country{USA}
}
\author{Sanorita Dey}
\email{sanorita@umbc.edu}
\orcid{0000-0003-3346-5886}
\affiliation{%
  \institution{University of Maryland, Baltimore County}
  \city{Baltimore}
  \state{Maryland}
  \country{USA}
}


\begin{abstract}
\bc{Organizations use asynchronous AI interview systems to efficiently manage large applicant pools, enabling quick and uniform evaluations. However, concerns remain about their impact on user agency and the lack of personalization applicants experience with these systems. Although efforts have been made to humanize the interview process, users’ expectations are often unmet, especially when compared to the promises made by these systems. To examine how applicants perceive and experience these tools, particularly in the context of their growing familiarity with large language models (LLMs), we conducted a two-phase study. The first phase involved an analysis of 11 subreddit discussions on interview experiences with asynchronous AI interviewers, followed by a semi-structured interview study with 17 participants. Qualitative analysis revealed key issues such as mismatched expectations, amplified by organizational rhetoric and applicant expectations shaped by experiences with LLMs. These factors shaped participants’ sense of agency and trust, often leading to workarounds and deceptive practices. In the follow-up study, we designed an interface with two features, response variants and feedback variants, and evaluated it across six groups (N = 180, 30 participants each) to assess whether these features support users’ sense of agency, competence, and relatedness. Our analysis suggests that even subtle design changes can enhance user autonomy and that carefully designed feedback can provide meaningful support in high-stakes interview contexts.}
\end{abstract}




\maketitle

\newcommand{\redtext}[1]{\textcolor{red}{#1}}
\newcommand{\bluetext}[1]{\textcolor{blue}{#1}}

\section{Introduction}

The adoption of Artificial Intelligence (AI) in recruitment represents a major transformation in human resource management (HRM), aimed at increasing efficiency and fairness in candidate evaluation. AI systems now perform functions traditionally handled by recruiters, including parsing resumes, analyzing candidate responses, and interpreting multimodal data such as text, facial cues, and voice \cite{gurdian_AI}. Applications span semantic matching \cite{zhang2018resumevis}, interview analysis \cite{qin2023automatic}, performance and attrition prediction \cite{trinkenreich2024predicting}, fair ranking \cite{geyik2019fairness}, gamified assessments \cite{jia2016personality}, and biometric screening \cite{reicherts2022s}. While AI-driven technologies offer speed and scalability, they raise concerns around bias, limited contextual understanding, and a lack of transparency \cite{biehl2025prestige}. In response to rising user expectations, many platforms now integrate emotional AI and Large Language Models (LLMs) to create more interactive interview experiences, both synchronous and asynchronous \cite{roemmich2023values, de2023performing}. However, the shift toward AI-driven evaluations has raised critical questions about fairness, candidate autonomy, and the erosion of human judgment in high stakes decision-making.

Building on these concerns, the CSCW community has critically explored the social and ethical dimensions of AI in recruitment. Emotion AI has been a key focus, with studies linking applicant perceptions to various forms of injustice \cite{pyle2024us}, revealing inflated claims about solving hiring “inaccuracy” and “inauthenticity” \cite{roemmich2023values}, and exposing ethical risks in workplace patents that disproportionately affect low-power workers \cite{boyd2023automated}. Emotional surveillance has also been shown to erode trust and wellbeing \cite{andalibi2024should, corvite2023data}. Social class bias persists, as hiring practices often favor communication styles associated with upper-middle-class backgrounds \cite{chua2022substance, chua2020you}, while referrals remain more effective than algorithmic assessments \cite{armstrong2025navigating}. Despite interest in fairness, practitioners struggle to apply formal fairness metrics in practice \cite{sarkar2024s}. 

Recent research has shown that AI-mediated interviews shape not only candidate behavior but also perceptions of fairness, control, and authenticity. In asynchronous formats, automated evaluations tend to reduce deceptive behavior, yet they also restrict candidates’ ability to express themselves fully, making the interaction feel constrained and impersonal \cite{suen2025comparing}. Many applicants perceive algorithmic hiring as less fair than human or hybrid processes, even when they receive favorable outcomes, due to a lack of recognition of their individuality and context \cite{lavanchy2023applicants}. These concerns can arise even before the interview begins; job advertisements that mention AI-led interviews often reduce organizational appeal and applicants’ intention to apply \cite{wesche2021repelled}. This discomfort becomes more pronounced in fully automated, high stakes settings, where candidates report lower feelings of control, fairness, and social presence compared to human-led formats \cite{langer2019highly}. Although AI is often associated with efficiency and objectivity, many candidates still prefer human involvement for its capacity to show empathy and adapt to context \cite{mirowska2022preferring}. Design elements such as warm avatars and informative feedback have been shown to improve perceptions of fairness and interpersonal respect \cite{min2024avatar}. On a behavioral level, AI evaluation has been found to improve speech rate and uncertainty, suggesting that candidates adjust their communication style in response to reduced social cues and feedback \cite{liu2023speech}. While asynchronous interviews generally lead to less deceptive impression management than synchronous ones, both AI and human raters struggle to detect such behaviors accurately \cite{suen2025comparing}. The use of AI-driven bots as interviewers in virtual environments also introduces concerns around trust, authenticity, and emotional connection \cite{hasler2013virtual}. Moreover, scholars have questioned whether AI assessments can adequately account for socially constructed traits, raising concerns about how identity and intent are constrained in such systems \cite{aizenberg2025examining}.

While prior work has shed light on candidate perceptions and design features of AI-driven interviews, important gaps remain. Most studies rely on controlled settings, offering limited insight into how job seekers engage with these systems in real-world contexts. Moreover, few examined how expectations have shifted with the emergence of LLMs. With the aim to address these gaps, we ask the following questions:

\begin{itemize}
\item \textbf{RQ1:} \oc{How do candidates experience and perceive asynchronous AI-driven interviews in the current landscape of widely accessible AI technologies? What factors contribute to the gap between the intended benefits of these AI-driven systems and the experiences reported by interviewees?}

\item \textbf{RQ2:} How might design considerations, informed by challenges identified in candidate experiences with AI-driven interviews, improve perceptions of control, support, and overall interaction quality?
\end{itemize}

To answer these questions, we conducted two studies. \oc{For RQ1, we analyzed Reddit discourse to capture unfiltered perspectives, followed by in-depth interviews to understand how candidates navigate, adapt to, and make sense of AI driven interviews in the context of widely accessible AI technologies}. We analyzed $\sim$18K Reddit data collected from 11 subreddits, related to applicants' experience of interacting with AI-driven interviewers. Afterwards, we conducted a semi-structured interview study with 17 participants. The interview study allowed us to go deeper into the challenges and concerns that we found from the Reddit analysis. Our qualitative analysis revealed that applicants often entered AI driven interviews with high expectations shaped by language or terminologies used by organizations and public familiarity with LLMs, only to be disappointed by conversational yet impersonal and rigid experiences. Some used workarounds to bypass the system, citing perceived unfairness. Neurodivergent individuals reported exclusion and shared coping strategies. While consistency and efficiency were acknowledged, acceptance was largely driven by reluctance rather than trust. The lack of transparency from hiring companies, absence of acknowledgment, and overall undesired experiences left participants feeling devalued, undermining both their sense of autonomy and self-efficacy.

\oc{Based on insights from the first study, which indicated that applicants often felt devalued due to a lack of personalized acknowledgment and perceived limited control over how they presented themselves, we explored two design considerations. They focused on providing personalized feedback and increasing applicants’ flexibility in responding.} We subsequently designed an interview interface incorporating two feature types, each offered in three variants: response variants (RV), which included only re-record, only edit, or both options; and feedback variants (FV), which included motivational, evaluative, or combined feedback. We evaluated these variants using the lens of Self-Determination Theory to examine how they might influence users’ sense of agency, competence, and relatedness, and in turn their performative confidence. We conducted a user study across six groups (N = 180, 30 participants each) and analyzed their post-task survey responses using a mixed-method approach. Our results suggest that response editing improved autonomy and competence, while re-recording was useful but sometimes frustrating, and combining both options introduced decision friction. Motivational feedback felt supportive, evaluative feedback was helpful but distant, and their combination sometimes lacked clarity, highlighting the need for thoughtful and flexible design.

The main contributions of our work are as follows:
\begin{itemize}
    \item Presenting large-scale analysis of 11 subreddit discourse with AI-driven asynchronous interviews to uncover how job seekers experience and adapt to AI-driven interviews in real-world contexts.
    \item Identifying key experiential gaps arising from organizational language, familiarity with contemporary AI technologies, and system limitations, with particular attention to how these affect autonomy, trust, and inclusion.
    \item Designing an AI-driven interview interface and evaluating it through the lens of Self-Determination Theory to examine how response and feedback features can improve user experience by supporting agency, competence, and emotional engagement.
    \item Extending the CSCW research direction on the adoption of new technologies by focusing on how people’s perceptions of social and organizational factors influence both technology adoption and the development of critical workarounds.
\end{itemize}

We organize the paper as follows. In Section \ref{relatedwork}, we review related work. Section \ref{method_1} and Section \ref{result_1} describe the methodology and findings of Study 1, while Section \ref{method_2} and Section \ref{result_2} present the method and results of Study 2. We conclude by discussing the broader implications of our work for the CSCW community, AI mediated hiring practices, and the design of human-centered AI (HCAI) systems.

\section{Related Work}
\label{relatedwork}

\subsection{Reframing Interviews in the Age of AI}
\subsubsection{From Dyadic Interaction to Algorithmic Assessment}
The interview process has long been a subject of HCI research, particularly concerning how relational dynamics shape candidate experiences~\cite{tullar1989relational}. Human interviews involve co-regulated verbal and nonverbal cues, with applicants expected to demonstrate both technical and interpersonal skills~\cite{ford2017tech}. However, applicants often misjudge their performance, struggle to manage anxiety-induced behaviors, and fail to mask unconscious nonverbal cues~\cite{sieverding2009cool, dipboye2012selection}. AI-driven interviews, especially in asynchronous settings, have fundamentally altered this interaction. Applicants must now perform without real-time feedback or mutual adaptation, leading to heightened uncertainty, faster speech, and more constrained responses \cite{lavanchy2023applicants, aizenberg2025examining}.

Many AI interview platforms now incorporate emotion AI to assess facial expressions, voice tone, and behavioral micro-features ~\cite{roemmich2023emotion, nawaz2020artificial}. These assessments are often based on invalidated or pseudo-scientific assumptions \cite{roemmich2023values}, and their use has raised critical concerns about reinforcing biases and penalizing marginalized candidates for behaviors outside dominant affective norms~\cite{mujtaba2019ethical}. Studies show that such systems can undermine applicant autonomy, especially when candidates cannot contest or understand the evaluation process \cite{andalibi2024should, pyle2024us, corvite2023data}, leaving applicants feeling dehumanized by opaque systems that offer little room for context or individuality. Particularly for neurodivergent or culturally diverse users, emotion-based evaluations risk flattening complex identities and failing to accommodate variation in expressive behavior \cite{chua2020you, liu2023speech}.

\subsubsection{Perceptions of Fairness, Trust, and Transparency}
Despite claims of improved efficiency and fairness, AI hiring systems often evoke skepticism and distrust. Applicants express concern over algorithmic opacity~\cite{balasubramaniam2023transparency, chua2022substance}, limited legal accountability~\cite{buyl2022tackling}, and the absence of explainability or recourse mechanisms \cite{fabris2025fairness, boyd2023automated}. While some users perceive AI as less biased than humans~\cite{kim2022artificial}, many remain uncomfortable with high-stakes decisions made without human oversight. Hybrid formats that combine AI assessments with human involvement are generally preferred \cite{aizenberg2025examining} especially for roles requiring empathy or contextual understanding \cite{mirowska2022preferring}. Notably, even including the word ``AI'' in job postings can lower candidate interest, suggesting deep-seated apprehension about algorithmic coldness and lack of fairness \cite{park2021human, wesche2021repelled}. Recent advances such as LLM-based interviewers aim to simulate more natural interaction~\cite{li2025metaagents, gan2024application}, but they raise similar concerns regarding trust, output reliability, and how well they represent human values~\cite{park2022designing, liu2023speech}. Applicants still feel the lack of transparency on how their inputs are processed, interpreted, and scored.

\subsubsection{Behavioral Responses and Design Directions}
AI interviews influence candidate behavior in distinct ways: participants tend to speak less, show fewer expressions, and reduce impression management, largely due to stress, uncertainty, and lack of reciprocal cues \cite{wesche2021repelled, langer2020anybody, liu2023speech}. While interface enhancements such as warm avatars, motivational prompts, or expressive feedback may improve perceptions of fairness and reduce anxiety~\cite{gonzalez2022allying, kim2022artificial, min2024avatar}, these adjustments often remain cosmetic. Without transparency, interpretability, or space for meaningful self-representation, candidates continue to face emotional labor, disempowerment, and limited control over how they are evaluated.

As platforms like HireVue\footnote{\url{www.hirevue.com/}}, Quinncia\footnote{\url{https://quinncia.io/}}, and other asynchronous tools became widely adopted, researchers advocate for candidate-centered approaches, emphasizing transparency, contestability, procedural fairness, and support for individual voice~\cite{li2021algorithmic, sumathi2024ai, hunkenschroer2023ai}. However, many AI systems still neglect context-sensitive traits such as resilience or lived experience, reinforcing behavioral norms that privilege dominant cultural expressions while marginalizing applicants from diverse socio-economic and cultural backgrounds~\cite{mujtaba2019ethical, roemmich2023emotion, chua2020you}.

\subsection{Supporting Agency and Engagement through Design}
Self-Determination Theory (SDT) has become a foundational lens in HCI for understanding how systems can support users’ basic psychological needs: autonomy, competence, and relatedness. In the context of conversational agents, SDT-informed design emphasizes personalization, flexible dialogue, and control over data to support autonomy and competence, though effectively providing relatedness remains challenging \cite{yang2021designing}. Broader critiques argue that SDT is often applied superficially; particularly in gaming and educational technologies where it serves more as a design heuristic than a robust theoretical framework \cite{ballou2022self, tyack2024self, dhiman2024designing}. Empirical research has shown that SDT-informed features, such as setting clear goals and incorporating perspective, enhance engagement and learning in instructional systems, and that positive user experiences can arise even from extrinsically motivated use, depending on how systems frame interactions \cite{bennett2024beyond}. However, attempts to establish connection through design may unintentionally compromise autonomy when relational dynamics are not carefully managed \cite{wenhart2024there}.

In proactive voice assistants, users respond more positively to systems that clearly communicate intent and respect user choice, underscoring the need for transparent, autonomy-supportive interactions even in anticipatory technologies \cite{hwang2022ai}. AI-mediated feedback environments also demonstrate that the framing and tone of feedback significantly affect user engagement. Narrative empathy, co-constructed responses, and the presence of a perceived partnership all improve feedback acceptance and sustain motivation over time \cite{yang2025control, nguyen2020challenges, oh2024better}. 

\oc{In applied domains such as hiring, health, and education, studies emphasize the importance of aligning system automation with user expectations \cite{li2021algorithmic, jacobs2021designing, pozdniakov2022question}}. Employers often prioritize technical over interpersonal skills, revealing a potential disconnect with AI-driven assessments \cite{kim2019co}. Across learning and review platforms, users benefit most when systems scaffold feedback exchanges, offer structure, and allow for agency-preserving participation \cite{adnin2022hiring, imam2024determining, cheng2020critique}. Whether in peer feedback, educational critique, or idea evaluation, the tone, timing, and transparency of system responses significantly influence user motivation, reinforcing the value of dialogic and autonomy-supportive designs \cite{foong2017online, dillahunt2020positive, yang2025understanding}.

This study employs SDT as an evaluative framework to examine whether applicants feel supported in their needs for autonomy, competence, and relatedness during AI interviews. By testing different variants of response and feedback features, we explore how the interfaces affect applicants’ sense of control, competence, and support throughout the process.

\section{Study 1: Methodology}
\label{method_1}
\begin{figure*}[t]
    \centering
    \includegraphics[width=1\linewidth]{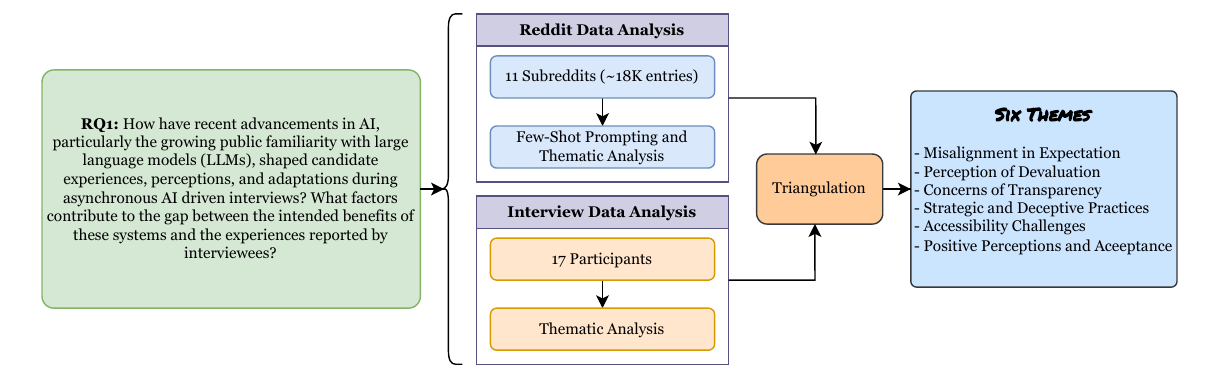}
    \caption{Flow-diagram of Study-1 aiming to answer RQ1 through a qualitative analysis on \textit{Reddit} discussion and interview data}
    \label{fig:study1}
\end{figure*}

Study 1 aimed to examine applicants’ perceptions and expectations of AI-driven asynchronous interviewers, particularly in the context of increasing access to LLM-based systems. We began by analyzing \textit{Reddit} data, followed by a semi-structured interview study. To ensure a diverse range of perspectives, we selected discussions from several relevant subreddits. Based on the insights gained, we conducted in-depth interviews to further explore emerging themes. The following section outlines the methodology of Study 1 in detail (shown in Figure \ref{fig:study1}).

\subsection{\textit{Reddit} Data Analysis}
To explore the public discussion around AI-driven interviews and automated hiring, we analyzed \textit{Reddit} data across communities where users discuss recruitment journeys, interview experiences, and algorithmic hiring systems. We initially identified a pool of subreddits based on prior literature \cite{stillman2023measuring, garg2021using, grenier2024automated} and curated lists from job-related websites \cite{sub_list1, sub_list2}. We then applied a structured filtering strategy to shortlist 11 subreddits that met the following criteria: (1) a minimum of 20,000 members to ensure visibility and community engagement, (2) thematic relevance to experience regarding AI-driven interviewers assessed through post content and flair usage, (3) sufficient discussion volume on AI-driven interviews (e.g., at least 50 relevant posts), and (4) public accessibility to allow ethical data collection. The final set of subreddits included \textit{r/jobs, r/recruiting, r/recruitinghell, r/interviews, r/getemployed, r/antiwork, r/askHR, r/careerguidance, r/careers, r/careeradvice, and r/hiring}.

\begin{table*}[ht]
\centering
\footnotesize
\setlength{\tabcolsep}{2pt}
\renewcommand{\arraystretch}{1.1} 
\begin{tabular}{|l|
m{2.3cm}|m{1.5cm}|m{2cm}|
m{2.2cm}|m{2.2cm}|m{2.2cm}|}
\hline
\textbf{Subreddit} &
\makecell[tl]{Strategic Self-\\Presentation} &
\makecell[tl]{Perceived \\Disrespect} &
\makecell[tl]{Transparency} &
\makecell[tl]{Expectation \\ Misalignment} &
\makecell[tl]{Positive \\ Acceptance} &
\makecell[tl]{Accessibility \\ Challenges} \\
\hline
\texttt{r/jobs} & 1209 (32.66\%) & 715 (19.32\%) & 651 (17.59\%) & 467 (12.62\%) & 562 (15.18\%) & 97 (2.63\%) \\

\texttt{r/recruiting} & 299 (34.62\%) & 123 (14.26\%) & 227 (26.34\%) & 74 (8.60\%) & 138 (16.04\%) & 2 (<1.00\%) \\

\texttt{r/recruitinghell} & 696 (21.50\%) & 1221 (37.74\%) & 857 (26.48\%) & 178 (5.50\%) & 185 (5.73\%) & 98 (3.06\%) \\

\texttt{r/interviews} & 234 (37.95\%) & 143 (23.10\%) & 63 (10.15\%) & 108 (17.47\%) & 67 (10.85\%) & 2 (<1.00\%) \\

\texttt{r/getemployed} & 348 (31.51\%) & 215 (19.46\%) & 173 (15.71\%) & 223 (20.24\%) & 142 (12.82\%) & 3 (<1.00\%) \\

\texttt{r/antiwork} & 281 (10.61\%) & 1154 (43.51\%) & 490 (18.46\%) & 363 (13.70\%) & 240 (9.06\%) & 124 (4.67\%) \\

\texttt{r/AskHR} & 227 (16.69\%) & 363 (26.68\%) & 289 (21.19\%) & 189 (13.89\%) & 271 (19.90\%) & 23 (1.65\%) \\

\texttt{r/careerguidance} & 312 (13.31\%) & 317 (13.56\%) & 195 (8.34\%) & 739 (31.58\%) & 597 (25.50\%) & 181 (7.71\%) \\

\texttt{r/careers} & 38 (24.57\%) & 34 (21.68\%) & 20 (13.29\%) & 37 (23.80\%) & 26 (16.60\%) & 0 (<1.00\%) \\

\texttt{r/careeradvice} & 644 (34.38\%) & 337 (17.97\%) & 203 (10.83\%) & 386 (20.63\%) & 233 (12.44\%) & 70 (3.75\%) \\

\texttt{r/hiring} & 98 (25.60\%) & 38 (9.98\%) & 111 (29.03\%) & 58 (15.25\%) & 75 (19.65\%) & 2 (<1.00\%) \\

\hline
\end{tabular}
\caption{Distribution of the analyzed subreddit posts, comments and replies across different thematic areas.}
\label{tab:thematic_dist}
\end{table*}

Using the PRAW API, we collected \textit{Reddit} data from January 2023 to March 2025, applying a keyword filter derived from terminologies used in prior research to extract content related to AI-driven interviews such as ``AI interviewer'', ``Algorithmic hiring'', ``automation in recruitment''; full list in Appendix \ref{keywords}. This timeframe was selected to capture discussions following the introduction and widespread adoption of LLMs at the end of 2022. While some posts may reference interviews that occurred before 2023, their inclusion reflects how users chose to share or revisit those experiences during the period of increased public attention to LLMs. The initial dataset contained  25,154 entries. After manual review revealed many irrelevant entries, we refined the dataset using OpenAI’s GPT-4o model (temperature = 0.5) with few-shot prompting to identify entries describing experiences with AI-based asynchronous interviews, including justifications for relevance (shown in Appendix \ref{prompting}) \cite{yan2024human, xiao2023supporting}. \oc{The final dataset comprised ~18,285 entries, including 9,707 unique users, 597 posts, 12,727 comments, and 4,961 replies.} The average word counts were 225.4 (±305.1) for posts, 54.2 (±82.3) for comments, and 46.3 (±70.1) for replies. \oc{A Reddit post initiates a discussion thread, a comment is a response to the post, and a reply is a response to another comment within that thread.}

\bc{Three human coders conducted the \textit{Reddit} analysis using a coding reliability approach, following the procedures outlined by Braun and Clarke \cite{braun2006using, braun2019reflecting}. We adopted a data-driven, grounded theory framework to allow codes and themes to emerge from the data rather than imposing predefined categories. During the coding process, they (1) reviewed the dataset filtered by the LLM to verify both the inclusion criteria and the rationale for filtering, and (2) conducted a open coding analysis. \oc{For each subreddit, entries were first ordered by posting time from earliest to latest. We then drew an initial random sample of 100 entries to establish a baseline for code relevance and clarity. Our objective was to find themes from the posts and comments. Identifying the trend of discussion was out of scope of this paper. Thus, we decided to do random sampling from our dataset and intentionally avoided temporal data analysis. After this calibration step, thematic coding was conducted using randomly selected entries within each subreddit and continued iteratively until thematic saturation was reached, defined as the point at which no new codes emerged from additional data \cite{saunders2018saturation}.} \oc{In total, we analyzed 1,893 entries (782 unique users, ~197 posts, ~1,696 comments and replies), and each coder judged over 93\% of the LLM-selected data as relevant, supporting the appropriateness of the sampled content.} Once the initial codes were identified, three coders met for several iterations to discuss their individually identified codes and through axial coding they grouped the related codes to form a refined and converged coding framework. \oc{Finally, a selective coding method was applied to identify the core themes of discussion of the \textit{Reddit} communities on AI-driven.} Based on the final list of core themes, three coders independently coded an additional set of 1534 entries (including posts, comments, and replies). To assess inter-coder reliability, we calculated Fleiss' Kappa coefficient, which yielded a value of 0.84, indicating strong agreement among coders \cite{landis1977measurement}. The resulting themes represent the primary concerns and strategies shared by users across multiple subreddits. The distribution of themes across these subreddits is shown in Table \ref{tab:thematic_dist}.}

\begin{table*}[ht]
\centering
\footnotesize
\renewcommand{\arraystretch}{1.3}
\setlength{\tabcolsep}{6pt}
\begin{tabular}{|c|c|c|c|p{3.6cm}|c|c|}
\hline
\textbf{ID} & \textbf{Gender} & \textbf{Ethnicity} & \textbf{Status} & \textbf{Major / Field} & \textbf{Neurotype} & \textbf{\# AI Int.} \\
\hline
P1  & Male        & Asian        & Grad (PhD)     & Computer Science             & Neurotypical            & 3 \\
P2  & Female      & White        & Undergrad      & Psychology                   & Neurodivergent (ADHD)   & 2 \\
P3  & Male        & Black        & Working        & Business                     & Neurotypical            & 3 \\
P4  & Male        & Asian        & Grad (MSc)     & Data Science                 & Neurotypical            & 4 \\
P5  & Non-binary  & Hispanic     & Working        & UX Design                    & Neurotypical            & 2 \\
P6  & Female      & Asian        & Grad (PhD)     & Education Technology         & Neurotypical            & 3 \\
P7  & Male        & White        & Undergrad      & Sociology                    & Neurotypical            & 2 \\
P8  & Male        & Asian        & Grad (MSc)     & Computer Science             & Neurotypical            & 6 \\
P9  & Male        & Black        & Grad (MSc)     & Software Engineering                  & Neurotypical            & 3 \\
P10 & Female      & Asian        & Undergrad      & Computer Science             & Neurotypical            & 2 \\
P11 & Male        & Multiracial  & Grad (PhD)     & Political Science            & Neurotypical            & 3 \\
P12 & Male        & White        & Grad (MSc)     & Mechanical Engineering       & Neurotypical            & 3 \\
P13 & Non-binary  & White        & Grad (PhD)     & Human-Centered Computing     & Neurodivergent (ADHD)   & 2 \\
P14 & Female      & White        & Working        & Communication Studies        & Neurotypical            & 5 \\
P15 & Male        & Black        & Undergrad      & Computer Science             & Neurotypical            & 2 \\
P16 & Male        & White        & Undergrad      & Computer Engineering         & Neurotypical            & 2 \\
P17 & Male        & Asian        & Grad (PhD)     & Software Engineering         & Neurotypical            & 7 \\
\hline
\end{tabular}
\caption{Interview participants' demographics including gender, ethnicity, academic/work status, field, neurotype, and number of AI-driven interviews faced.}
\label{tab:participant_demographics}
\end{table*}




\subsection{Interview Data Analysis}

\bc{Reddit analysis allowed us to develop our primary understanding of the challenges experienced by interviewees by analyzing a diverse, large-scale dataset. This approach was critical in developing a broad range of perspectives, especially from people who might be difficult to reach in formal study settings. However, analyzing posts and comments of Reddit users did not allow us to ask follow-up questions which is essential to uncover clarifying intent, context, underlying motivations, and underrepresented or absent nuances. To fill these gaps, we conducted a follow-up interview study.}

Based on our \textit{Reddit} analysis, we formulated a questionnaire for an in-depth interview study. Following the themes that emerged from the \textit{Reddit} findings, we developed a semi-structured interview guide. The interview questions probed deeper to gather participants' experiences with AI-driven asynchronous interviews, focusing on areas such as perceived fairness, accessibility challenges, expectations versus reality, system transparency, data privacy concerns, and coping strategies used during the interview process (please refer Appendix \ref{int_qs} for the interview questions).

We invited participants from 12 U.S. universities using emails to student communities, alumni groups, and snowball sampling. Eligible participants were required to be at least 18 years old, proficient in English, and to have used an AI driven asynchronous interview system at least once. Before recruiting participants, we collected basic demographic information, including whether they identified as neurodivergent, as a significant portion of Reddit users with neurodiversity shared experiences related to AI interviews in these subreddits. We also asked preliminary questions about participants’ AI interview experiences, including how many times they had participated and how recently their most recent interview took place. We initially received 51 responses and prioritized participants who had interacted with AI driven interviewers in 2024 or later. To maintain diversity and achieve thematic saturation without oversampling \cite{saunders2018saturation}, participants were invited in batches of five, with recruitment continuing only when new or significant insights emerged. This process resulted in 17 completed interviews, with demographic details reported in Table \ref{tab:participant_demographics}. The interview study received IRB approval from the first authors’ institution. Interviews were conducted online via Google Meet by the first and second authors and lasted 38 minutes on average. In total, we analyzed approximately 10.8 hours of interview recordings, resulting in about 90,000 words of transcribed text. Each participant received a \$20 Amazon Gift Card.


\bc{We employed grounded theory (the same approach followed for analyzing the Reddit dataset) to analyze the interview data. The first and second authors independently conducted open coding following grounded theory principles \cite{thornberg2014grounded, khandkar2009open} for a subset of participants (N = 5), allowing conceptual codes to emerge inductively rather than imposing codes from the semi-structured interview guide. Through regular discussions and iterative refinements (axial coding), they consolidated these codes into a more organized coding structure \cite{cohen2017coding, kendall1999axial}. Finally, they identified the overall list of themes emerged from the interviews using selective coding. \oc{These themes were developed to capture the presence and range of issues discussed in the interviews, rather than to assess narrative depth or relative importance across data sources.} Two coders independently coded the interviews of the remaining participants (N = 12) based on the identified list of themes. To check consistency, the inter-rater reliability was calculated using Cohen's Kappa, which yielded a score of 0.87, indicating strong agreement. \oc{To enhance the robustness of our findings, we applied triangulation \cite{flick2004triangulation}, using interview data to clarify and contextualize themes identified in the \textit{Reddit} analysis rather than to introduce new dominant categories.}}

\subsection{Consolidation of Reddit and Interview Themes}

\oc{We used a convergent triangulation approach to thematically integrate the themes that emerged from the \textit{Reddit} and interview datasets, consistent with prior qualitative research \cite{carter2014use, gupta2025being, subramanian2021detectables, pimenova2025good}.} \bc{We iteratively examined the alignment between themes identified in the Reddit data and those observed in the interview data, without treating either source as a primary standard. This process allowed us to identify areas of convergence (themes present in both sources), complementarity (themes that elaborated on each other), and divergence (themes unique to one source). The themes supported by both data sets were merged and those unique to one data set were retained as such, with their origin explicitly indicated in Table \ref{tab:themes_subthemes_sources}. Throughout the analysis, we maintained source tags (\textit{Reddit} vs. interview) for all coded excerpts, which allowed us to trace how evidence from each dataset contributed to each sub-theme.}

\section{Study 1: Result}
\label{result_1}
In the following section, we present the key themes identified from our  \textit{Reddit} and interview data. Interview participants are labeled as ‘P’, and Redditors as ‘R’, with  \textit{Reddit} comments numbered sequentially for clarity. \oc{Most of the Redditors quoted in this section are unique users (25 out of 28).} For ethical purpose, we have paraphrased all the quotes we collected from  \textit{Reddit} without removing their essence and meaning. Table \ref{tab:themes_subthemes_sources} presents all the main themes identified in our analysis, along with their corresponding subthemes and associated data sources (interviews, \textit{Reddit}, or both).
\begin{table}[h!]
\centering
\footnotesize
\renewcommand{\arraystretch}{1.2} 
\setlength{\tabcolsep}{6pt} 
\begin{tabular}{|p{5.2cm}|p{5cm}|p{3cm}|}
\hline
\textbf{Main Theme} & \textbf{Sub-themes} & \textbf{Source} \\
\hline

\multirow{4}{*}{\colorbox{LightBlue}{\parbox{5cm}{\textbf{Misalignment Between Applicant Expectations}}}}
& Expectation of Human-Like Presence (Anthropomorphism) & Both \textit{Reddit} and Interview \\
\cline{2-3}
& Discrepancy Between Promotional Language and Actual Functionality & Both \textit{Reddit} and Interview \\
\cline{2-3}
& Expectation of LLM-like Capabilities & Both \textit{Reddit} and Interview\\
\cline{2-3}
& Difficulty for Non-Native Speakers Due to Accent Handling & Interview Only \\
\hline

\multirow{3}{*}{\colorbox{LightGreen}{\parbox{5cm}{\textbf{Perception of Disrespect and Devaluation}}}}
& Disrespect Stemming from One-Way, Non-Interactive Format & Both \textit{Reddit} and Interview \\
\cline{2-3}
& Dissatisfaction Directed at Hiring Process, Not AI Itself & Interview Only\\
\cline{2-3}
& Superficial Encouragement Without Genuine Responsiveness & Both \textit{Reddit} and Interview \\
\hline

\multirow{4}{*}{\colorbox{LightYellow}{\parbox{5cm}{\textbf{Concerns Regarding Transparency}}}}
& Insufficient Communication from Hiring Organizations (External Transparency) & Interview Only\\
\cline{2-3}
& Uncertainty About System Functioning and Evaluation Criteria (Internal Transparency) & Both \textit{Reddit} and Interview \\
\cline{2-3}
& Data Privacy and Security Concerns & Both \textit{Reddit} and Interview \\
\cline{2-3}
& Perceived Exploitation of Applicant Data & \textit{Reddit} Only \\
\hline

\multirow{3}{*}{\colorbox{LightPink}{\parbox{5cm}{\textbf{Strategic Representation and Deceptive Practices}}}}
& Strategic Performance and Impression Management & Both \textit{Reddit} and Interview \\
\cline{2-3}
& Leveraging External Tools to Bypass System Constraints & Both \textit{Reddit} and Interview \\
\cline{2-3}
& Moral Conflict and Justification Under Pressure & Both \textit{Reddit} and Interview \\
\hline

\multirow{3}{*}{\colorbox{LightLavender}{\parbox{5cm}{\textbf{Accessibility Challenges for Neurodivergent Candidate}}}}
& Risk of Misinterpreting Neurodivergent Behaviors & Both \textit{Reddit} and Interview \\
\cline{2-3}
& Reluctance to Disclose and Seek Accommodations & Interview Only \\
\cline{2-3}
& Emotional Strain from Lack of Human Feedback & Both \textit{Reddit} and Interview \\
\hline

\multirow{3}{*}{\colorbox{LightPeach}{\parbox{5cm}{\textbf{Positive Perception and Acceptance}}}}
& Convenience and Flexibility & Both \textit{Reddit} and Interview \\
\cline{2-3}
& Perceived Standardization and Fairness & Both \textit{Reddit} and Interview \\
\cline{2-3}
& Pragmatic Adaptation to an Inevitable System & Both \textit{Reddit} and Interview \\
\hline

\end{tabular}
\caption{Thematic Analysis of Study 1, presenting main themes, their sub-themes, and associated data source(s). Color codes are used to distinguish each main theme, and the same colors are applied to their corresponding sub-themes to maintain visual consistency when they are discussed later.}
\label{tab:themes_subthemes_sources}
\end{table}

\begin{table*}[ht]
\centering
\footnotesize
\renewcommand\theadfont{\bfseries}
\setlength{\tabcolsep}{6pt}
\renewcommand{\arraystretch}{1.5}
\begin{tabular}{|>{\centering\arraybackslash}m{1.2cm}|m{1.8cm}|m{10.8cm}|}
\hline

\multirow{9}{*}{
  \begin{minipage}[c][\dimexpr9\baselineskip][c]{\linewidth}
    \centering
    \vspace{50pt}  
    \rotatebox[origin=c]{90}{
      \shortstack{
        \textbf{\small\colorbox{LightBlue}{\textcolor{black}{Misalignment Between}}}\\
        \textbf{\small\colorbox{LightBlue}{\textcolor{black}{Applicant Expectations}}}
      }
    }
  \end{minipage}
}
& \makecell{\textbf{\small Participant/}\\\textbf{\small Redditor ID}} & \textbf{Quotes} \\
\cline{2-3}
& \textbf{P2} & As they said, I thought it’d be smart, like actually "do" something. But nope, just a blank screen and questions. Not even a personalized feedback. \\
\cline{2-3}
& \textbf{P3} & The whole thing felt like a marketing pitch gone wrong. They sold it like it was high-tech, but it didn’t even try to respond or adapt. \\
\cline{2-3}
& \textbf{P7} & If they just said it was pre-recorded, fine. But calling it ‘AI’ made me expect a conversation. There’s a responsibility in how you frame tools. \\
\cline{2-3}
& \textbf{R1} & ...Not sure if anything has changed. I worked with P\&G years ago and their AI interview tool looked fancy but gave zero context. I expected today's tools to interact at least a little, it didn’t... \\
\cline{2-3}
& \textbf{R2} & ...They said 'AI-powered,' I thought cool, like ChatGPT or something. Turns out it was just a glorified pre-recorded form. Disappointing. \\
\cline{2-3}
& \textbf{R3} & If they just told us upfront it was static, I’d manage my expectations. But calling it ‘conversational’ set me up for a tech letdown. I mean, just be clear next time. \\
\hline
\end{tabular}
\caption{Interview participants and Redditors' paraphrased quotes highlighting expectation mismatches in AI-driven interview experiences.}
\label{tab:theme1_quotes}
\end{table*}

\subsection{Misalignment Between Applicant Expectations and AI-Driven Interview Processes}

The mismatch of expectations emerged as one of the most prominent themes in our study. A substantial number of \textit{Reddit} comments and interview responses (number of \textit{Reddit} entries or NR= 66, number of interview participants or NI=8)  pointed to elements of \colorbox{LightBlue}{anthropomorphism}, such as remarks like “the system does not feel like a human” or “I cannot see the nonverbal cues during the interview,” which have been explored in prior research \cite{cohn2024believing, ivarsson2023suspicious, bi2023create}. However, our findings revealed two unique types of expectation discrepancies: (1) applicants observed a noticeable gap between the way interview tools were described and how they actually functioned (as pointed out by P3 and R2), and (2) applicants anticipated more advanced, conversational, and interactive systems that resembled contemporary AI technologies rather than the impersonal and static tools they encountered (shown in Table \ref{tab:theme1_quotes}).

Several interview participants (NI=5) noted that the instruction emails sent by the hiring organizations used phrases such as “AI-based,” “with our AI tool,” or “our conversational platform,” (an issue raised by P7 and R3) which led them to expect a technologically advanced system. In some cases, participants were provided with a practice link intended to help them become familiar with the platform. However, when they accessed the actual interview, the system appeared noticeably \colorbox{LightBlue}{different- both in design and interaction style} and often felt more like a static asynchronous video recording tool than the advanced technology they had anticipated (echoing concerns from P2 and R1). These findings reiterate that terminologies can create inflated expectations, and when unmet, lead to disappointment and distrust of the AI-systems (a reaction mirrored by R3 and P3) \cite{langer2022look}. As we heard from P2 said, 

\begin{quote}
\textit{``[...] they (hiring organization) said their AI recruiter would take the interview, but honestly, the system didn’t live up to that. The only AI thing about it was some avatar, and even that didn’t do much. You just answer the questions, and that’s it. I was expecting something more conversational, like how ChatGPT talks back, you know? After a while, it became hard to keep up the energy and interest. There was no back and forth, no real interaction, nothing that felt intelligent. It just started with a fixed set of questions, and no matter what I said, some generic text would pop up on the corner. It just blindly followed the list. I was hoping for something that could actually adapt to what I was saying.''}
\end{quote}

P2 added that the evaluation might have happened in the background by an AI-based algorithm but knowing it beforehand would have been better for them to prepare mentally what to expect on the spot. We noticed applicants (NR=31, NI=6) gradually shifted their expectations towards a more conversation like experience similar to their interactions with LLM-driven chatbots (e.g., ChatGPT, Gemini etc) as evident in our findings where R2 explicitly \colorbox{LightBlue}{compared their expectations to ``ChatGPT-like'' interaction} and R4 was expecting the system would build on their answers like an LLM does. R5 commented where the OP (original poster) shared an experience with an AI-driven interviewer:

\begin{quote}
\textit{``[...] there needs to be better communication or use of this tool. As the job market has become tighter, people already feel devalued when they are interviewed by an automated system. And it gets worse when it is not even up to the standard. Basically, you are just talking to a box at best and being profiled at worst."- R5}
\end{quote}

Apart from the lack of conversational features, non-native English-speaking participants (NI=3) shared their struggles with systems that poorly transcribed their responses. Prior experience with LLMs shaped expectations of comparable support from the interviewer. However, the experience was otherwise, as the transcription errors affected how their answers were represented, leading to frustration and a sense of being misjudged. When asked how they knew the system was not using an LLM to capture their response, P1 answered:"

\begin{quote}
    \textit{``I know based on past experiences with Siri and Alexa. These systems tend to struggle with non-major English accents. That's why I try to enunciate with an American accent when speaking to these tools. But I saw that ChatGPT can capture my sentences even when I am speaking with my natural accent. That's how I know these tools (AI hiring tools) are not using LLM features.''- P1}
\end{quote}

Another participant, P8, added that \colorbox{LightBlue}{masking an accent} is itself a stressful task. But knowing that their accent could be the biggest barrier for them to secure their desired job made the experience even more frustrating. They shared:

\begin{quote}
\textit{“You know you are already at a disadvantage because English is not your first language, and I admit I should be more fluent in speaking. But when you are in an interview and the experience makes you feel excluded, even when the technology exists to do better, it just makes you feel small.”- P8}
\end{quote}

\subsection{Perceptions of Devaluation in One-Sided or Non-Interactive Interview Formats}

\begin{table*}[ht]
\centering
\footnotesize
\renewcommand\theadfont{\bfseries}
\setlength{\tabcolsep}{6pt}
\renewcommand{\arraystretch}{1.5}
\begin{tabular}{|>{\centering\arraybackslash}m{1.2cm}|m{1.8cm}|m{10.8cm}|}
\hline

\multirow{9}{*}{
  \begin{minipage}[c][\dimexpr9\baselineskip][c]{\linewidth}
    \centering
    \vspace{60pt}  
    \rotatebox[origin=c]{90}{
      \shortstack{
        \textbf{\colorbox{LightGreen}{Perceptions of Disrespect and Devaluation}}
      }
    }
  \end{minipage}
}
& \makecell{\textbf{\small Participant/}\\\textbf{\small Redditor ID}} & \textbf{Quotes} \\
\cline{2-3}
& \textbf{P8} & If it’s gonna replace a person, it needs to at least feel present. This felt like a surveillance camera with a mic \\
\cline{2-3}
& \textbf{P11} & ...It wasn’t even the tech that bothered me, it was the complete lack of acknowledgment. Like, I said all that just to get a generic ‘thank you for your submission’?... \\
\cline{2-3}
& \textbf{P10} & It’s not even about being anti-AI. I’d take a smart bot over a rude recruiter any day. But this? This was just me performing to silence. No response, no presence, just me and my doubts. \\
\cline{2-3}
& \textbf{P15} & It so feels like a cog in the wheel. You are definitely not valued as a human, just another datapoint in their pipeline. \\
\cline{2-3}
\cline{2-3}
& \textbf{R6} & ...The system said ‘you’re doing great’ after every question. Sweet, but who decided that? It felt less like reassurance and more like automated flattery... \\
\cline{2-3}
& \textbf{R7} & ...If this is the future of hiring, count me out. I don’t need a robot to smile at me, but at least pretend like my response mattered... \\
\cline{2-3}
& \textbf{R8} & ...There is no way in freaking hell I will ever do another one-way video interview. It’s awkward, unnatural, and makes my anxiety spike through the roof. At least with a real person, I can read the room.... \\
\hline
\end{tabular}
\caption{Interview participants and Redditors' paraphrased quotes highlighting perceptions of disrespect and devaluation in one-sided interview formats.}
\label{tab:theme2_quotes}
\end{table*}

A recurring theme we found was the applicants’ \colorbox{LightGreen}{perception of disrespect and devaluation} within the one-sided, asynchronous AI-driven interview format. Prior research has similarly documented resistance toward AI technologies when they are perceived as replacements for human roles \cite{woodruff2024knowledge, thakkar2020towards}. Although many comments on \textit{Reddit} (NR=42) expressed strong aversion toward AI-driven interviewer tools, we observed a number of participants from the interview study are not entirely opposed to AI-driven interviewers themselves. Rather, their \colorbox{LightGreen}{dissatisfaction is more directed toward the broader hiring process} and the rigid, non-interactive interview format as also described by P10 calling it ``performing to silence'' (shown in Table \ref{tab:theme2_quotes}). While various design considerations have been proposed to improve the asynchronous interview experience, this gap in user acceptance and perceived fairness remains unresolved. As R9 commented on  \textit{Reddit},

\begin{quote}
    \textit{``I have done numerous interviews with AI.[..] so it's not that people are resisting the use of AI in the hiring process, they're just resisting the disrespectful model of the one-way video interview.''- R9}
\end{quote}

This tension highlights a deeper issue: advancements in AI, have not translated into a more human-centered interview experience. Despite improvements in interface design and system responsiveness, applicants continue to feel unseen and devalued suggesting that the core problem lies not in technical capability (as reported by P10 and P11), but in the lack of reciprocal, relational interaction. P15 mentioned that the recruitment tool he faced used LLM as per the description in their website. The overall system felt very advanced, and the interface was highly intuitive, but they still felt devalued by the one-sided nature of the interview:

\begin{quote}
    \textit{``There might’ve been an LLM behind the screen, but to me, it still felt like no one was listening.''- P15}
\end{quote}

Another participant, P16, shared a similar experience, noting that the system was smooth and the supportive text pop-ups, such as brief prompts like \textit{``Take your time''} or \textit{``You are doing great''} helped create a more comfortable environment. However, they still felt unacknowledged during the interaction, as their responses \colorbox{LightGreen}{received no apparent recognition} which reinforced the sense of a one-sided exchange (as R6 described it as a hollow reassurance). This impression was further reinforced when the final feedback appeared overly generic, suggesting that their answers had not been individually considered.

\begin{quote}
    \textit{``[...] It (the final feedback) all looked polished, but it felt like the outcome was decided before I even spoke, more like an automated summary than a response to what I actually said.''- P16}
\end{quote}

\subsection{Concerns Regarding Transparency and Algorithmic Decision-Making}

\begin{table*}[ht]
\centering
\footnotesize
\renewcommand\theadfont{\bfseries}
\setlength{\tabcolsep}{6pt}
\renewcommand{\arraystretch}{1.5}
\begin{tabular}{|>{\centering\arraybackslash}m{1.2cm}|m{1.8cm}|m{10.8cm}|}
\hline

\multirow{9}{*}{
  \begin{minipage}[c][\dimexpr9\baselineskip][c]{\linewidth}
    \centering
    \vspace{60pt}  
    \rotatebox[origin=c]{90}{
      \shortstack{
        \textbf{\colorbox{LightYellow}{Concerns Regarding Transparency}}
      }
    }
  \end{minipage}
}
& \makecell{\textbf{\small Participant/}\\\textbf{\small Redditor ID}} & \textbf{Quotes} \\
\cline{2-3}
& \textbf{P1} & I wasn’t asking them to explain the entire algorithm. But just a simple breakdown of what’s being evaluated, like whether it’s just verbal response or also non-verbal cues would’ve helped me prepare more confidently. \\
\cline{2-3}
& \textbf{P4} & Even after completing the process, I wasn’t sure if a human ever saw my video. It’s hard to reflect or improve when you don’t even know who or what was reviewing your performance. \\
\cline{2-3}
& \textbf{P9} & I think transparency could actually improve trust. A little more clarity around what’s analyzed and how data is handled would go a long way toward making the experience feel more ethical and respectful. \\
\cline{2-3}
& \textbf{R10} & ...They call it ‘interview automation’ but it’s just secret surveillance wrapped in startup buzzwords. If you’re gonna judge me for eye contact, at least say so... \\
\cline{2-3}
& \textbf{R11} & ...Not HireVue, but I did one where I had to answer timed questions on camera. I kept thinking ‘If I pause to think, will it look like I’m unprepared?’ My anxiety made me rush, and I hate how that probably affected how I came off... \\
\cline{2-3}
& \textbf{R12} &  ...Like others have said, interviews are a two way street. At the very least, tell me what you’re scoring or looking for. If it’s all automated, that’s fine but let me know. Otherwise, it just feels manipulative. \\
\cline{2-3}
& \textbf{R13} & ...Probably free data for their AI, not a real interview. They just want a wide range of responses to train their algorithm. No real interest in hiring.... \\

\hline
\end{tabular}
\caption{Interview participants and Redditors' paraphrased quotes highlighting concerns regarding transparency and algorithmic decision-making}
\label{tab:theme3_quotes}
\end{table*}

Transparency has been a major topic of discussion for black box systems. There are many concerns reported about the AI-driven interviewer in previous studies, which we also observed in our study \cite{andalibi2024should, stark2021ethics, armstrong2023navigating}. However, our findings focused more on the transparency related to both external (communication from hiring organizations) and internal (functioning of the AI-driven systems) aspects of the interview journey. We learned from interviewees that applicants are \colorbox{LightYellow}{often less informed or not informed at all} about critical interview-related details. Although it is understandable that hiring organizations may not reveal everything about their hiring process, as per the participants, organizations failed to provide "necessary details" to make the experience better as echoed by P1, who wished for even a basic outline of what would be evaluated, including non-verbal cues (shown in Table \ref{tab:theme3_quotes}).

Participants described gaps in both external and internal transparency that shaped their experience of the interview. Several interview participants (NI=6) noted that, in some cases, they were not given clear guidelines on how to approach the interview. This pointed to a gap in \textit{external transparency}, as organizations provided only general instructions about logistics, such as checking the internet connection or using a computer, without clarifying the interview format or what would be evaluated. At the same time, participants (NR=39, NI=8) reported a lack of \textit{internal transparency} about how the AI-driven systems operated. There was little to no information about \colorbox{LightYellow}{what to expect from the system itself}. Although the tools appeared advanced in terms of interface design, applicants were unsure whether behaviors such as facial expressions, eye contact, or body language would be assessed. R10 described this uncertainty as ``secret surveillance wrapped in startup buzzwords.'' This confusion was also reflected in P17’s experience, who shared that before their first AI-driven interview, the organization sent an email that referred only to a ``standard interview procedure,'' which gave basic setup tips but offered no explanation of how the system functioned or what it would assess.

\begin{quote}
    \textit{``I was really confused. The good part was that the system looked impressive, but since it was my first attempt, and there were no proper guidelines, it felt like I was taking a test without being told the rules. I had no idea if I was being recorded for an actual evaluation or just for show. Was the system monitoring my facial expressions? Should I keep eye contact all the time or look away occasionally? Should I smile while answering? Should I wear formal attire? All these questions continued to run through my mind. And that kind of guessing game really messes with your confidence.''- P17}
\end{quote}

Another major concern revealed in our findings was the data privacy issues. Participants (NR=78, NI=7)\colorbox{LightYellow}{voiced} \colorbox{LightYellow}{significant privacy concerns} regarding their recorded responses. While hiring organizations captured their video interviews, they were not clearly informed about how these recordings would be stored, who would have access to those recordings, or how long the data would be retained. This lack of transparency raised fears about potential misuse, unauthorized sharing, or data breaches. R14 commented on a class action lawsuit in which a major pharmaceutical company is sued for secretly using AI technology to analyze facial expressions during interviews and assigning candidates an 'employability score' without informing them \cite{cvs_lawsuit},

\begin{quote}
    \textit{``After that whole CVS and HireVue mess where they secretly tracked people's facial expressions and scored them without telling anyone, it is no wonder nobody trusts AI interviewer agents anymore.''- R14}  
\end{quote}

This \colorbox{LightYellow}{fear of data exploitation} was also shaped by other experiences. Redditors (NR=13) mentioned feeling disrespected by the one-sided format of AI-driven interviews, facing rejection without explanation, and often being ghosted after completing the process \cite{lyons2024vanishing}. These experiences led some to conclude that the organizations were using the interview primarily as a means to harvest data rather than to truly consider them for the position (as R13 described “free data for their AI”). As shared by one of the Redditors, R15:

\begin{quote}
    \textit{``It felt less like a job interview and more like unpaid data entry for their algorithm.''- R15}  
\end{quote}

Our study revealed concerns about the lack of transparency in how AI systems make decisions, particularly in evaluating interpersonal behaviors. One common example was eye contact. Participants were told to maintain eye contact during interviews but were not informed how the system interpreted it. Many looked away briefly to think and later got worried that this might be interpreted as disengagement. Moreover, it was unclear whether a human would later review their responses or if the evaluation was entirely automated (as mentioned by P4). This uncertainty left participants unsure whether to use specific keywords to appease an algorithm or speak naturally to connect with a possible human reviewer.

\subsection{Strategic Self-Presentation and Deceptive Practices in AI Interview Responses}

\begin{table*}[ht]
\centering
\footnotesize
\renewcommand\theadfont{\bfseries}
\setlength{\tabcolsep}{6pt}
\renewcommand{\arraystretch}{1.5}
\begin{tabular}{|>{\centering\arraybackslash}m{1.2cm}|m{1.8cm}|m{10.8cm}|}
\hline

\multirow{9}{*}{
  \begin{minipage}[c][\dimexpr9\baselineskip][c]{\linewidth}
    \centering
    \vspace{60pt}  
    \rotatebox[origin=c]{90}{
      \shortstack{
        \textbf{\colorbox{LightPink}{Strategic Self-Presentation}}\\
        \textbf{\colorbox{LightPink}{and Deceptive Practices}}
      }
    }
  \end{minipage}
}
& \makecell{\textbf{\small Participant/}\\\textbf{\small Redditor ID}} & \textbf{Quotes} \\
\cline{2-3}
& \textbf{P3} & Of course you can game the system. You can game anything once you figure out how it scores. AI interviews are just another test. You learn the rules, then play to win. \\
\cline{2-3}
& \textbf{P12} & I filtered my answers. Not because I wanted to, but because I felt like the system couldn’t handle nuance \\

\cline{2-3}
& \textbf{R16} & ...AI interview has generic questions like, ‘Tell me about a time you worked on a team.’ It’s less about being honest and more about hitting the right phrases. You start answering like you're writing SEO copy... \\
\cline{2-3}
& \textbf{R17} & ...So how exactly does an AI interview do that? You just trick the system by giving overly structured, polished answers. It's not about who you are, it’s about how well you play the game. It's not even real... \\
\cline{2-3}
& \textbf{R18} & ...I tried it out to see what response I’d get. And it felt like the whole thing encouraged me to rehearse, not reflect. The moment I tried to speak naturally, I worried I’d be penalized for going off-script.... \\
\cline{2-3}
& \textbf{R19} & ...Do you mind if I use your video but instead of your voice, I dub in an AI-generated voice and see if I get more hits? Just testing the system's biases... \\
\hline
\end{tabular}
\caption{Interview participants and Redditors' paraphrased quotes highlighting concerns regarding strategic self-presentation and deceptive practices in AI interview responses}
\label{tab:theme4_quotes}
\end{table*}
From  \textit{Reddit}, we collected numerous comments (NR=167) that largely focused on how to navigate AI-driven interviews and the deceptive practices that some applicants use to get through them (shown in Table \ref{tab:theme4_quotes}). Since our selected subreddits were about tips and tricks to do well in the interviews, many Redditors (e.g., R16, R17, R18) shared their experience to help others but at the same time we found people mentioning about deceptive strategies which they adopted often in time of desperation. 

As part of their \colorbox{LightPink}{self-presentation strategies}, several interview participants and  \textit{Reddit} users (NR=41, NI=9) described intentionally using certain keywords during the interview. This aligns with prior findings noting that Applicant Tracking Systems (ATS) often scan for such keywords as part of automated evaluation processes \cite{cappelli2019your}. \oc{Some of the users (NR=18, NI=5) reported exaggerating positivity, energy, or expressions (e.g., forced smiles or sustained eye contact) in response to assumed non-verbal tracking by the system, a tactic known as deceptive impression management (IM) by faking impression also in prior research \cite{suen2025comparing}.} Some shared to memorize pre-written responses with correct pronunciation to common questions to appear more polished to create positive impressions. One of the participants in our interview study (P14) described preparing for the interview “like a hackathon” emphasizing the strategic nature of performance under constrained interaction.

\begin{quote}
    \textit{``I planned it like a hackathon. Calm face, solid voice, keywords on cue. Can’t risk surprises with a system that can’t ask follow-ups.''- P14}  
\end{quote}

This kind of performative action was seen to be exhausting. For example, R20 humorously captured this pressure by saying:

\begin{quote}
    \textit{``At this point, I’m not sure if I’m applying for a job or auditioning for an Oscar. Smiled like a maniac, nailed the keywords, and prayed that the AI liked my lighting setup. Still got ghosted''- R20}  
\end{quote}

\textit{Reddit} users (NR=26) often \colorbox{LightPink}{described using external aids to game up} the interview process. Many were frustrated by the lack of non-verbal cues and the limit on re-recordings, which made it hard to refine their answers. In some cases, they had no retry option and had to do it in a single attempt. Therefore, they turned to LLMs to craft polished responses on the first try and even used AI voice bots to deliver them as described in a comment ``..just let AI talk to AI''. While we do not endorse such practices in any circumstances, they point to a broader issue: When systems feel rigid or opaque, applicants may look for workarounds. Some users \colorbox{LightPink}{expressed guilt, but justified their actions} due to high stakes, perceived unfairness, or a belief that the system was already biased. P3 shared that going through a one-sided interview with templated, inflexible response formats was particularly challenging for them. They added that, in such situations, they could understand why someone might feel compelled to bend the rules or seek outside help just to get through the process.

\begin{quote}
    \textit{``I’ve heard of people using ChatGPT or notes to get through these interviews. I never did it myself, but I can understand why someone would, especially if they felt they weren’t being given a fair shot. Sometimes it’s not about cheating, it’s about surviving a system that already feels stacked against you.''- P3}  
\end{quote}

\subsection{Accessibility Challenges for Neurodivergent Candidates in AI-driven Interviews }

\begin{table*}[ht]
\centering
\footnotesize
\renewcommand\theadfont{\bfseries}
\setlength{\tabcolsep}{6pt}
\renewcommand{\arraystretch}{1.5}
\begin{tabular}{|>{\centering\arraybackslash}m{1.2cm}|m{1.8cm}|m{10.8cm}|}
\hline

\multirow{9}{*}{
  \begin{minipage}[c][\dimexpr9\baselineskip][c]{\linewidth}
    \centering
    \vspace{60pt}  
    \rotatebox[origin=c]{90}{
      \shortstack{
        \textbf{\colorbox{LightLavender}{Accessibility Challenges}}\\
        \textbf{\colorbox{LightLavender}{for Neurodivergent Candidates}}
      }
    }
  \end{minipage}
}
& \makecell{\textbf{\small Participant/}\\\textbf{\small Redditor ID}} & \textbf{Quotes} \\
\cline{2-3}
& \textbf{P2} & I really tried to stay focused, but staring at the camera with no feedback just made my brain wander. With ADHD, I rely a lot on cues, tone, facial cues, little reactions - to stay anchored. This was just silence. By the third question, I was already mentally checked out. \textbf{(ADHD)} \\
\cline{2-3}
& \textbf{P13} & I spent more time prepping how to act than how to answer. Should I pause more? Smile? Look at the lens the whole time? ADHD makes me second-guess everything in that kind of setting. And because there was no feedback, I kept overthinking it even after it ended. \textbf{(ADHD)} \\
\cline{2-3}
& \textbf{R21} & ...I had an interview where the company asked me to keep my eyes on the camera the whole time. I couldn’t focus on the question and think clearly while also staring at the lens. I kept wondering if looking away for a second would cost me... \textbf{(Autism)} \\
\cline{2-3}
& \textbf{R22} & ...What’s annoying is that when I didn’t use AI to help write my answers, I got ghosted. But when I used it to ‘polish’ my wording, I got callbacks. Makes you wonder how much authenticity actually matters or if it just rewards people who mask better... \textbf{(Autism)} \\
\cline{2-3}
& \textbf{R23} & ...Same thing happened to me... when they called, I disclosed that I have anxiety and ADHD. After that, I never heard back. Automated systems don’t care about that stuff.... \textbf{(ADHD)} \\
\cline{2-3}
\hline
\end{tabular}
\caption{Interview participants and Redditors' paraphrased quotes highlighting concerns regarding accessibility challenges for neurodivergent candidates in AI-driven interviews}
\label{tab:theme5_quotes}
\end{table*}

While prior research has examined accessibility and design concerns through structured interviews \cite{ara2024collaborative}, our analysis of unsolicited narratives from public forums such as  \textit{Reddit} suggests additional areas that may warrant attention. These include \colorbox{LightLavender}{possible misinterpretations of neurodivergent behaviors}, emotional challenges associated with systems lacking human feedback, and the ways in which some users may adapt strategically to such environments. While some users (NR=9, NI=2) shared coping strategies (see Table \ref{tab:theme5_quotes}), others reported consistently negative outcomes, particularly when AI systems appeared to evaluate non-verbal behaviors that may differ from neurotypical norms. One Redditor, R24, reflected:

\begin{quote}
    \textit{``I’m autistic and have never even landed a second interview after a Workday screening. I looked into it and found that the AI evaluates facial expressions, eye contact, and body language to predict job performance. It feels completely discriminatory.''- R24}  
\end{quote}

Despite companies offering adjustments, this support often depends on candidates self-identifying and requesting accommodations \cite{bbc-neuro}, which many are reluctant to do. This hesitation often leads neurodivergent applicants to \colorbox{LightLavender}{withhold their condition} and instead find alternative, less direct ways to manage the process. As P13 shared:

\begin{quote}
    \textit{``I no longer disclose my condition because the accommodations offered have been ineffective or have ended up working against me. Instead, I ask to spread long interview rounds over multiple days, but frame it as a scheduling issue rather than a disability-related need.''- P13}  
\end{quote}

Neurodivergent individuals often navigate social interactions by interpreting subtle contextual cues; intonation, pacing, facial feedback that help them adjust and respond appropriately \cite{zolyomi2024emotion}. In digital systems that lack these human elements, especially AI-driven interview platforms or automated assessment tools, this interpretive space collapses. The \colorbox{LightLavender}{absence of reassurance or clarification} can lead to heightened anxiety, self-doubt, and a persistent feeling of being misjudged or misunderstood. Rather than adapting to the system, users may find themselves overcompensating or disengaging entirely. These experiences highlight the emotional friction that arises when interfaces are designed without responsiveness or recognition of differences. One Redditor, R25, who is autistic, shared that they did not receive additional accommodations because the hiring organization did not have such a provision. They proceeded with the interview anyway, but found it to be a rather unpleasant experience:

\begin{quote}
    \textit{``It’s like talking into a mirror that doesn’t reflect anything back. I can’t tell if I’m being understood, or if I’ve made a mistake. So I rehearse every word, not because it helps, but because I’m scared [that] the system will see me as wrong just for being me.''- R25}  
\end{quote}

\subsection{Positive Perceptions and Acceptance of AI-Driven Interview Systems}

\begin{table*}[ht]
\centering
\footnotesize
\renewcommand\theadfont{\bfseries}
\setlength{\tabcolsep}{6pt}
\renewcommand{\arraystretch}{1.5}
\begin{tabular}{|>{\centering\arraybackslash}m{1.2cm}|m{1.8cm}|m{10.8cm}|}
\hline

\multirow{9}{*}{
  \begin{minipage}[c][\dimexpr9\baselineskip][c]{\linewidth}
    \centering
    \vspace{40pt}  
    \rotatebox[origin=c]{90}{
      \shortstack{
        \textbf{\colorbox{LightPeach}{Positive Perception}}\\
        \textbf{\colorbox{LightPeach}{and Acceptance}}
      }
    }
  \end{minipage}
}
& \makecell{\textbf{\small Participant/}\\\textbf{\small Redditor ID}} & \textbf{Quotes} \\
\cline{2-3}
& \textbf{P4} & Honestly, I liked how structured it was. No small talk, no guesswork, just give your answer and move on. \\
\cline{2-3}
& \textbf{P17} & The process was fair in the sense that everyone got the same setup. No interviewer bias, no weird vibes, just consistency. \\
\cline{2-3}
& \textbf{R25} & ...Was it weird talking to my webcam? Yeah. But I did it at 2 AM in pajamas with my dog next to me. Try doing that in a regular interview... \\
\cline{2-3}
& \textbf{R26} & ...Say what you want about robots, but at least they don’t care if you forgot to iron your shirt or have a weird laugh. Everyone gets the same setup, and honestly, that feels fairer sometimes.. \\
\cline{2-3}
& \textbf{R27} & ...Look, AI interviews aren’t perfect, but neither is traffic or 9 AM calls. The tech will get better, but meanwhile, I’d rather learn how to ride the wave than wait for it to crash... \\
\cline{2-3}
\hline
\end{tabular}
\caption{Interview participants and Redditors' paraphrased quotes highlighting positive perception and acceptance of AI-driven interview systems}
\label{tab:theme6_quotes}
\end{table*}

Despite various flaws and limitations, both \textit{Reddit} comments and interview participants (NR=61, NI=10) highlighted several positive aspects of AI-driven interview systems (shown in Table \ref{tab:theme6_quotes}). Many Redditors \colorbox{LightPeach}{appreciated the} \colorbox{LightPeach}{speed, scheduling flexibility,} and automation these systems offer. Such features were seen to alleviate some of the logistical burdens typically associated with traditional interviews. P17 described the system as “convenient” when applying to multiple positions. In particular, one-way or pre-recorded formats enabled some users to prepare in a low-pressure environment and better manage social anxiety. As one Redditor, R28, shared:

\begin{quote}
\textit{``I prefer recording answers alone rather than speaking to a panel. It was better than being stared at by four people on Zoom.''- R28}
\end{quote}

Some Redditors viewed AI-based interviews as \colorbox{LightPeach}{more standardized or impartial,} suggesting that they reduce certain human biases and promote equal treatment. Similar views emerged in our interviews. P9 noted that while the system's uniformity ensures everyone faces the same environment, this can be both its strength and its limitation:


Others took a pragmatic stance, recognizing that AI-driven interviews, although imperfect, have become a routine part of hiring. Rather than dwelling on their flaws, participants \colorbox{LightPeach}{emphasized adapting and moving forward,} noting that earlier formats had their own limitations as well.

P9 further remarked that when system shortcomings are evident, hiring organizations have the responsibility to address them. A well-designed tool is not enough on its own; it must be paired with thoughtful implementation:

\begin{quote}
\textit{``The issue arises when the system is mediocre and the hiring organization is also ignorant about their responsibilities. Undoubtedly, they can be more sincere.'' – P9}
\end{quote}

Some users also reflected on how they learned to adapt to the system by understanding how AI evaluates responses. They argued that while the process may be uncomfortable, resisting it is less productive than learning how to navigate it, especially as it becomes more common in hiring.

\subsection{Summary}
Our findings reveal a disconnection between applicants’ expectations and their actual experiences with AI driven interviews. Many expectations were shaped by the language used by hiring organizations and the growing public familiarity with LLMs (e.g., ChatGPT, Gemini), which led some to assume that the systems would be highly responsive and personalized. Participants expressed disappointment when the interviews appeared conversational but lacked personalized interaction or adaptation. Several applicants described using strategies to trick or bypass the system, often justifying their actions by pointing to its rigidity and perceived lack of fairness. Neurodivergent individuals reported experiences of exclusion and shared specific workarounds they developed to cope with the process. While many described the experience as dehumanizing, some acknowledged benefits such as consistency and efficiency. Still, acceptance was often marked by resignation rather than trust, shaped by the absence of better alternatives.

\section{Study 2: Methodology}
\label{method_2}

\begin{figure*}[t]
    \centering
    \includegraphics[width=0.8\linewidth]{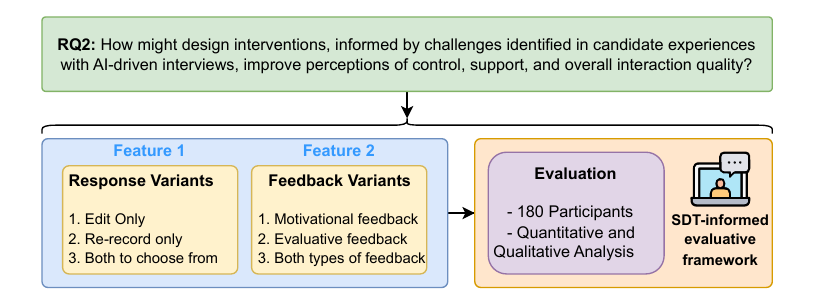}
    \caption{Flow-diagram of Study-2 aiming to answer RQ2 through different variants of two features separately; response type and feedback type followed by a mixed-method analysis}
    \label{fig:study2}
\end{figure*}

\bc{Building on the findings of Study 1, we conducted a follow-up study to explore design opportunities for improving applicant experience. Study 1 showed that while participants acknowledged the benefits of AI in recruitment, many participants were not entirely positive about AI-driven interviews. After becoming familiar with LLMs, their expectations had shifted, and they wanted interviews to feel more personalized and human-centered. They often felt devalued, seeing this as a result of organizational neglect or the tool’s limitations, and wanted more acknowledgment during their interactions with the AI-driven interview system. Guided by these insights, we considered features for an asynchronous AI-driven interviewer to facilitate a stronger sense of personalization and value. We refer to each design consideration as a \textit{feature} and its specific interface implementation as a \textit{variant}. The following section explains our design considerations in detail, followed by the interface implementations and the data analysis approach (see Figure \ref{fig:study2}).}

\subsection{Design Considerations and Motivation}

\bc{Our initial study showed that many applicants felt disrespected by the one-sided nature of AI interviews, citing a lack of acknowledgment and control over how they could present themselves. These concerns reflected broader issues of motivation, fairness, and self-representation. To address them, we focused on two key directions that consistently emerged across interviews and \textit{Reddit} discussions: enhancing applicants’ sense of agency and their sense of belonging. Based on these insights, we established the following design considerations (DCs):}

\subsubsection{\textbf{DC1:} Give interviewees more agency to ensure their responses are interpreted as intended}\leavevmode\\
\bc{Interview participants (e.g., P1, P2, P3, P7, P10, P12, P14, P17) and Redditors (e.g., R3, R5, R8, R14, R19) described feeling devalued during AI-driven interviews because they could not tell how the system interpreted their answers. Many noted that organizations gave little explanation or transparency about how responses were assessed, which made them feel disregarded. This uncertainty led them to second-guess their words, remove nuance, or mimic what they thought the system wanted, which reduced the authenticity of their self-presentation. Several reported over-rehearsing or changing their tone and accent out of fear of being misunderstood. With the human element removed from these high-stakes evaluations, they felt organizations had a greater obligation to ensure fairness and clarity. As one participant said, \textit{``I wasn’t asking them to explain the entire algorithm. But just a simple breakdown of what’s being evaluated… would’ve helped me prepare more confidently'' -P1}. These challenges show why giving interviewees more agency is vital. Supporting their sense of autonomy can enable authentic self-expression, reduce anxiety about misinterpretation, and help them present their abilities with confidence.}

\subsubsection{\textbf{DC2:} Give interviewees personalized feedback to help them feel acknowledged and valued}\leavevmode\\
\bc{Interview participants (e.g., P2, P8, P10, P11, P15, P16) and Redditors (e.g., R2, R6, R9, R13, R21) repeatedly noted that AI-driven interviews made them feel invisible because they received no acknowledgment or personalized response to their efforts. Many described putting substantial thought into their answers only to get generic confirmation messages, which made the process feel mechanical and dismissive. This lack of recognition eroded their sense of belonging and made them feel like ``just another entry in a dataset.'' One participant reflected, \textit{``...It wasn’t even the tech that bothered me, it was the complete lack of acknowledgment...” -P11}. A Redditor echoed, \textit{``It felt like yelling into a void. If they can analyze me, they can at least tell me something back'' -R6}. These reactions show why personalized feedback is essential: it would counteract feelings of devaluation, signal that their individual contributions matter, and help restore a sense of human connection in the interview process.}

\subsection{Interface Design and Implementation}
\begin{figure*}[t]
    \centering
    \includegraphics[width=1\linewidth]{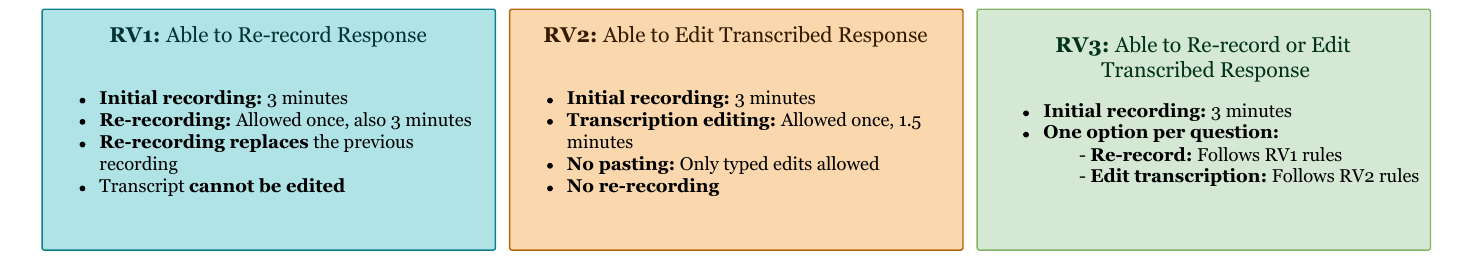}
     \caption{Design details of different variants of response interface: RV1 only lets re-record, RV2 only lets edit and RV3 provides both options to choose from}
    \label{fig:response}

\end{figure*}

For \textbf{DC\textsubscript{1}}, we envisioned several design choices and decided to include a feature that allows applicants to re-record their responses. Our goal was not only to display their recorded response as transcribed text but also to let them correct unintended errors. We designed three \textbf{response variants (RV)} for this interface. The first variant, \textit{Re-record only}, lets users record a new response if they find the transcription significantly inaccurate. Although several existing AI interview systems allow re-recording, we included this variant mainly as a baseline for comparison. The second variant, \textit{Edit only}, allows users to revise the transcribed text within a limited time window. This design was informed by participants who shared that they sometimes wanted to make small changes or add details but were hesitant to start over from scratch. The third variant, \textit{Both}, allows users to choose between re-recording and editing. We hypothesized that providing both options would enhance their sense of agency and reduce stress \cite{nielsen1994enhancing}. The design of these variants is shown in Figure \ref{fig:response}.

For \textbf{DC\textsubscript{2}}, our objective was to provide some form of personalized messages or responses to users so that they can feel valued. One way to achieve that goal is through personalized feedback.  Prior research suggests that such feedback can enhance user engagement and trust in AI systems \cite{han2025self, dowthwaite2025exploring}. After careful consideration, we designed another interface with three types of \textbf{feedback variants (FV)} to fulfill \textbf{DC\textsubscript{2}}. (1) motivational feedback, aimed at acknowledging participants' effort and maintaining engagement \cite{hassan2019motivational}; (2) evaluative feedback, offering real-time suggestions \cite{liu2025interview}; and (3) combined feedback, which included both types of feedback (see Appendix~\ref{prompt_feedback}). Motivational feedback was designed to reinforce reflection, and emotional safety, deliberately avoiding critique or correction. Evaluative feedback, in contrast, offered warm but targeted suggestions in-between the questions to help users refine their responses based on the STAR framework (Situation, Task, Action, Result) which is a common rubric to evaluate behavioral response \cite{star_method}. We avoided using any numerical scoring as we did not want to cause performance anxiety or comparison pressure. In the combined condition, both types were shown, with motivational feedback presented first to align with positive framing principles \cite{wu2021better}.

\begin{figure*}[t]
    \centering
    \includegraphics[width=1\linewidth]{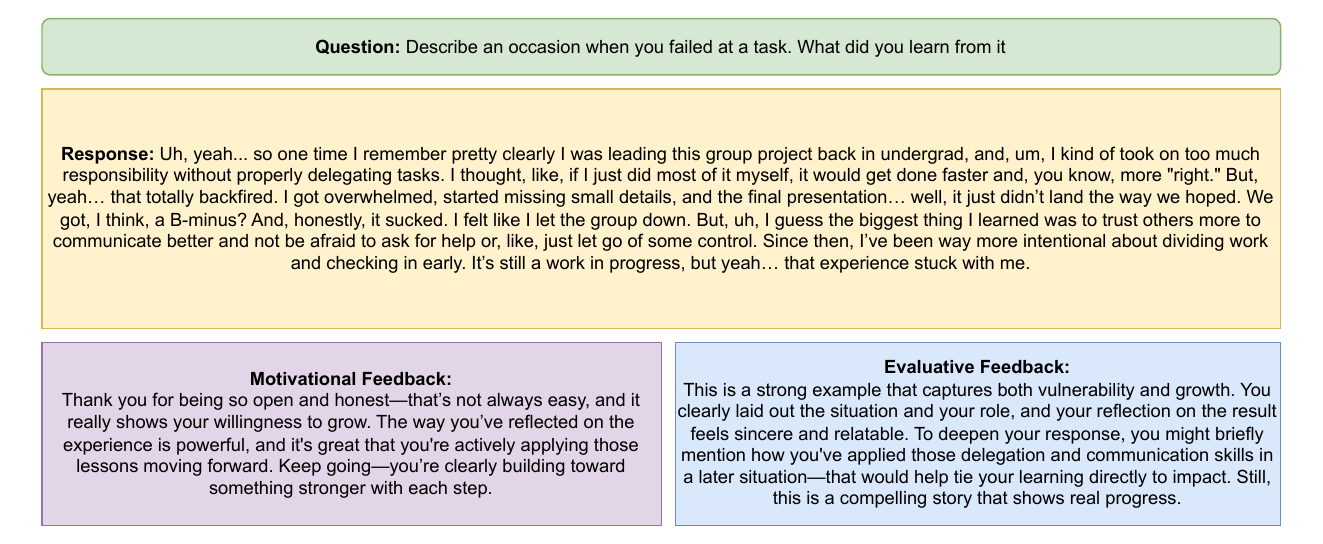}
     \caption{An example of motivational and evaluative feedback for a behavioral interview response to \textit{"Describe an occasion when you failed at a task. What did you learn from it?"}. The motivational feedback emphasizes encouragement and personal growth, while the evaluative feedback provides constructive suggestions grounded in the STAR (Situation, Task, Action, Result) framework.}
    \label{fig:feedback}

\end{figure*}  

\begin{figure*}[t]
    \centering
    \includegraphics[width=0.9\linewidth]{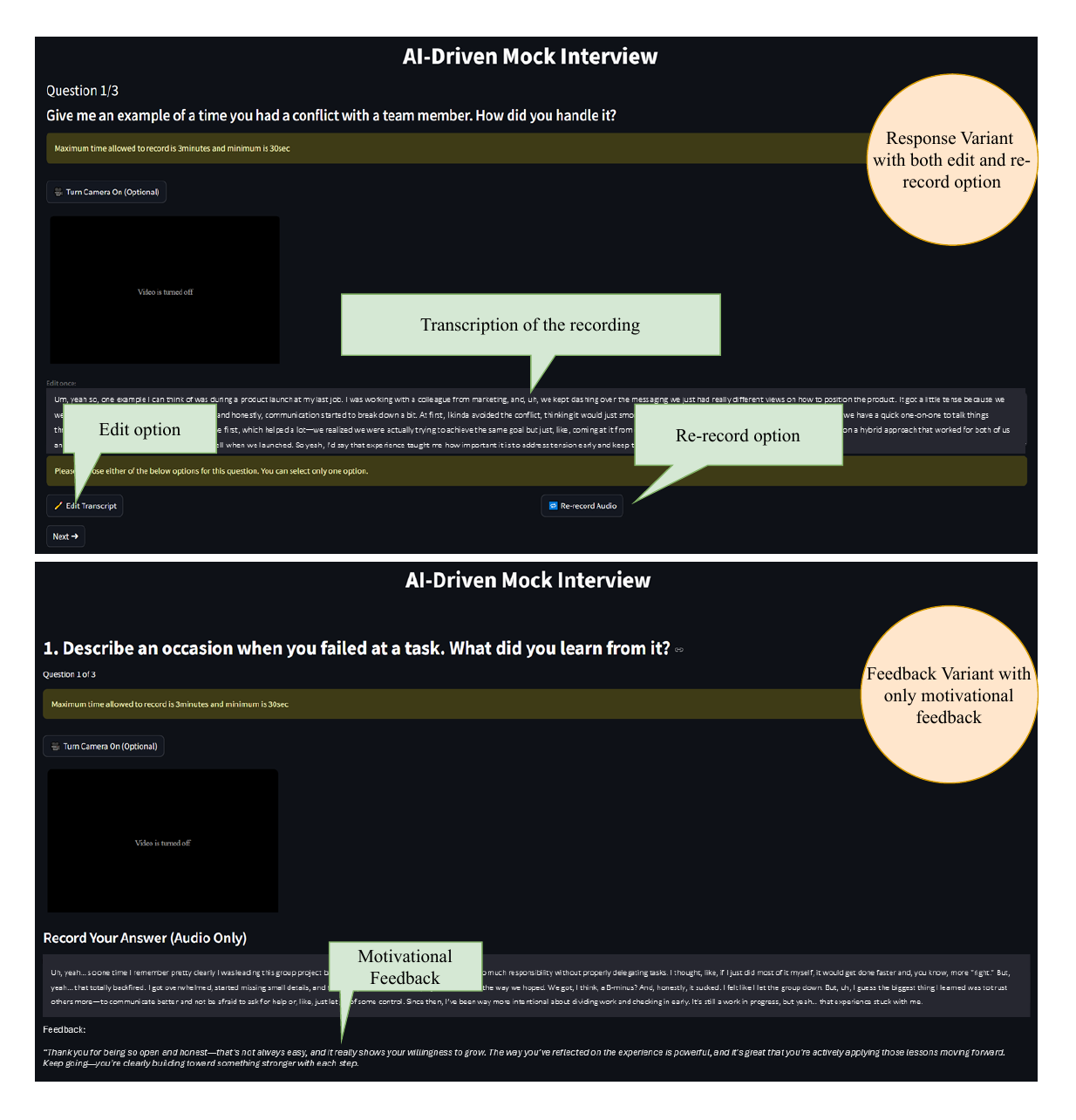}
    \caption{Snapshot of our system: Shows the response variant-3 or RV3, where the user can see both options (edit or re-record) after recording their response (above), and the feedback variant-1 or FV1, where the user can only get motivational feedback (below)}
    \label{fig:interface_ss}
\end{figure*}

For our study, we developed a web-based interface that simulated an AI-driven interviewer system using Streamlit\footnote{\url{https://streamlit.io/}}, customized with CSS to control the layout and visual design. To provide participants with a realistic interview experience, we instructed them to treat the session as a mock interview. Audio recordings were transcribed via OpenAI’s Whisper API\footnote{\url{https://openai.com/index/whisper/}}. To support our experimental design, we built separate interfaces for RVs and FVs. As these two features had three variants, we ended up building in total of six different variants. In the RVs, participants could either re-record or edit their response. But in the FVs, participants were not allowed to revise their responses. Instead, they viewed the transcription and received personalized feedback generated by GPT-4o through few-shot prompting (see Appendix \ref{prompt_feedback}). We recorded all transcriptions, feedback, and system activity logs for analysis. The interface was designed to be minimal and intuitive, supporting ease of use. The application was hosted online to enable remote participation (shown in Figure \ref{fig:interface_ss}).

\subsection{User Study and Data Analysis}

\subsubsection{Theoretical Framework and Measures}\leavevmode\\
To evaluate and compare the efficacy of each variant, we used \textit{Self-Determination Theory (SDT)} as a guiding framework, following prior HCI and CSCW studies that applied SDT to design and evaluate interactive systems \cite{ryan2000self, dhiman2024designing, yang2021designing, han2025self}. We chose SDT because our design considerations align with its three psychological needs: autonomy, competence, and relatedness. \textbf{DC\textsubscript{1}} (response variants or RVs) was expected to support autonomy by allowing users to control how their responses were captured, and competence by helping them express themselves more effectively. \textbf{DC\textsubscript{2}} (feedback variants or FVs) was intended to support relatedness by helping users feel acknowledged and emotionally supported, and to support competence by boosting their confidence through constructive guidance. 

We developed custom self-report items inspired by SDT-informed practices in HCI research. For the RVs, we asked participants if they felt a greater sense of control (autonomy), if they could express themselves more effectively (competence), if they felt confident in their self-presentation (competence), and if they found the feature useful (a pragmatic usability measure). For the FVs feature, we asked if they found the feedback useful (usability), if it increased their confidence in their competence (competence), if they felt acknowledged (relatedness), and if it enhanced their sense of engagement (motivation). While these items were not taken from a standardized SDT scale, they follow established SDT-based HCI practices that use task-specific measures aligned with autonomy, competence, and relatedness \cite{yang2021designing, hicks2016framing}.

\subsubsection{Participants}\leavevmode\\
We recruited 180 participants from 19 U.S. universities through student community emails, alumni networks, and snowball sampling. Eligibility criteria included being at least 18 years old, fluent in English, and having prior experience with AI-driven asynchronous interviews. Participants from Study 1 were excluded to minimize bias. Each participant received \$10 USD after completing the study. The study received approval from the authors’ Institutional Review Board (IRB) office. 

Participants ranged in age from 18 to 36 years. Of the 180, 102 identified as male, 64 as female, 10 as non-binary or another gender, and 4 preferred not to say. Ethnic backgrounds included Asian (74), White (54), Black or African American (34), Multiracial (10), and Other (8). Most participants were graduate students (92), followed by undergraduates (46) and employed individuals. All participants had prior experience with AI-driven interviews: 72 had used them once, 82 had done so 2–4 times, and 26 had used them more than four times.

\subsubsection{Study Design and Procedure}\leavevmode\\
\oc{We used a between-subjects design with six groups of 30 participants each. Participants were assigned to one of three response variant (RV) conditions (P1--P30; P31--P60; P61--P90) or one of three feedback variant (FV) conditions (P91--P120; P121--P150; P151--P180).} The response variants did not include any feedback, and the feedback variants did not allow users to re-record or edit their responses. This separation was intentional to avoid interaction effects and to isolate the individual impact of each feature.

\begin{figure*}[t]
    \centering
    \includegraphics[width=0.6\linewidth]{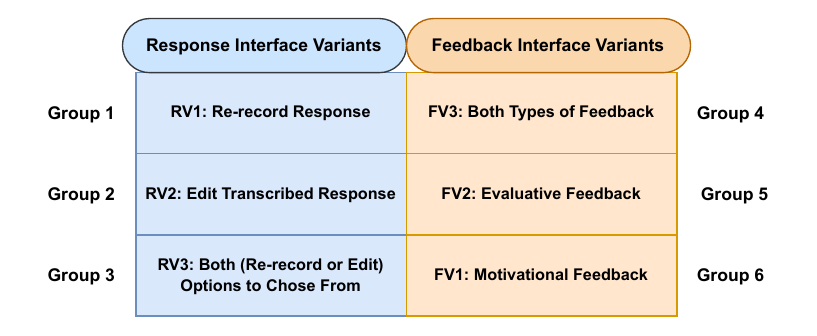}
    \caption{Group-wise allocation of different variants of response and feedback interfaces. (RVs = Response Variants, FVs = Feedback Variants)}
    \label{fig:grp_interface}
\end{figure*}

Participants were asked to approach the mock interview as if it were a real one, to simulate the sense of sincerity and engagement found in high-stakes settings. To prevent unnecessary stress, we informed them that their responses would not be evaluated. We also provided a briefing that explained what the system would collect, how their data would be used, and what to expect during the study. During the session, participants recorded responses to behavioral interview questions using our system (see Appendix \ref{behave_qs}) \cite{indeed_qs}. Depending on their assigned group, they encountered different interface layouts (Figure \ref{fig:grp_interface}). After completing the task, they filled out a post-task survey that included Likert scale items (1 = Strongly disagree, 5 = Strongly agree) and an open-ended question about their experience. The entire task was conducted remotely. 

\subsubsection{Data Analysis}\leavevmode\\
We conducted Welch ANOVA with Games-Howell post hoc tests to compare Likert-scale responses across interface variants. Two independent coders analyzed the open-ended responses thematically using a grounded theory approach. They began by familiarizing themselves with the data and independently generating initial codes through open coding. They then iteratively refined the codebook through axial coding, consolidating related codes and building a shared understanding of emerging concepts. Using selective coding, they identified the main themes across responses. After establishing the thematic structure, both coders independently applied the agreed-upon themes to the remaining responses. We assessed inter-rater reliability using Cohen’s kappa (RVs: 0.81, FVs: 0.84), which indicated a high level of agreement.

\section{Study 2: Result}
\label{result_2}

\subsection{Response Interface Variants: Quantitative Analysis}
We evaluated the effect of three response variants: RV1 (Re-record), RV2 (Edit), and RV3 (Both) on four outcome variables: sense of control, expressiveness, confidence, and usefulness. A Welch ANOVA was used to test for differences among groups. The results are presented in Table~\ref{tab:transcript}.

We found statistically significant differences for sense of control ($F(2,57) = 6.83$, $p = .001$, $\eta^2 = .047$), confidence ($F(2,56) = 3.65$, $p = .028$, $\eta^2 = .029$), and expressiveness ($F(2,57) = 6.11$, $p = .003$, $\eta^2 = .041$). According to the effect sizes, these differences are small to moderate, but meaningful in the context of interaction design.

Post hoc Games–Howell comparisons revealed that for sense of control, participants rated RV2 (Edit) significantly higher than both RV1 (Re-record) and RV3 (Both). For expressiveness, RV2 was again significantly more preferred than RV3, and also showed a modest advantage over RV1. In terms of confidence, RV3 outperformed RV1, but the difference between RV2 and RV3 was not significant.

No significant differences were observed for usefulness ($p = .075$) although descriptively, RV1 scored slightly higher. This suggests that while participants viewed all variants as generally fair and respectful, these dimensions did not vary strongly with interaction design.

\begin{table*}[ht]
\centering
\footnotesize
\renewcommand{\arraystretch}{1.3}
\setlength{\tabcolsep}{5pt}
\begin{tabular}{|l|c|c|c|p{3.6cm}|}
\hline
\textbf{Variable} & \textbf{RV1 (Re-record)} & \textbf{RV2 (Edit)} & \textbf{RV3 (Both Options)} & \textbf{Post-Hoc Analysis} \\
\hline
\makecell[l]{\textbf{Sense of Control} \\ $F(2,57) = 6.83$, \textbf{$p = .001$}} & 
\textbf{M = 3.567}, SD = 0.497 &
\textbf{M = 3.678}, SD = 0.470 &
M = 3.489, SD = 0.503 &
RV2 > RV3*; RV2 > RV1* \\
\hline
\makecell[l]{\textbf{Usefulness} \\ $F(2,57) = 2.62$, $p = .075$} & 
\textbf{M = 3.567}, SD = 0.498 &
M = 3.400, SD = 0.501 &
M = 3.456, SD = 0.493 &
— \\
\hline
\makecell[l]{\textbf{Confidence} \\ $F(2,56) = 3.65$, \textbf{$p = .028$}} & 
M = 3.322, SD = 0.577 &
M = 3.500, SD = 0.545 &
\textbf{M = 3.533}, SD = 0.503 &
RV3 > RV1* \\
\hline
\makecell[l]{\textbf{Expressiveness} \\ $F(2,57) = 6.11$, \textbf{$p = .003$}} & 
M = 3.211, SD = 0.743 &
\textbf{M = 3.322}, SD = 0.668 &
M = 2.967, SD = 0.694 &
RV2 > RV3**; RV1 > RV3* \\
\hline
\end{tabular}
\caption{Welch ANOVA and Games–Howell post hoc comparisons across response variants or RV. Significant post-hoc differences are marked with asterisks (* $p < .05$, ** $p < .01$, *** $p < .001$).}
\label{tab:transcript}
\end{table*}
\subsection{Response Interface Variants: Qualitative Analysis}
\subsubsection{Re-record Response (RV1): Simplicity with Repetition Cost} \leavevmode\\
Participants using RV1 often commented on its simplicity and predictability. There was no ambiguity in how to proceed: if something went wrong, the only option was to re-record. For some users, this structure was helpful and easy to follow. However, many found it mentally tiring and time-consuming. Small mistakes meant starting over entirely, leading to frustration and disengagement over time. Some participants reported accepting suboptimal recordings rather than starting again, indicating a trade-off between effort and satisfaction. \oc{Others felt discouraged by the high stakes attached to minor slip-ups, which reduced their sense of confidence as mentioned by P12, \textit{``Even tiny slip-ups felt costly because there was no way to fix just that part''}.} Although the interface was easy to understand, its lack of granularity seemed to amplify emotional pressure. Therefore, RV1 may provide procedural clarity but limits opportunities for low-effort recovery, which can gradually reduce user confidence with repeated use. This aligns with the lowest confidence score for RV1 and its lower sense of control compared to RV2.

\subsubsection{Edit Response (RV2): Clarity, Control, and Unexpected Pressure}\leavevmode\\
RV2 was most commonly described as precise, efficient, and satisfying. Participants found editing opportunity helpful for correcting minor errors without needing to repeat their full response. The process felt streamlined and low-effort, especially when compared to re-recording. Many comments reflected that this tool enabled users to make quick adjustments and feel in control of their final message. \oc{P33 shared, \textit{``It felt like I had more control over my response, which made the process satisfying.''}.} This is reflected in RV2’s top scores for both sense of control and self-representation. However, a smaller subset of participants used editing not only to correct errors but also to reword or reshape entire thoughts. These users engaged in multiple rounds of revision, not out of necessity, but from a concern about tone, impression, or misinterpretation. For them, editing prompted a type of performative caution, where self-presentation mattered as much as correctness. This behavior introduces a possible tension: while editing increased perceived control, it may also have increased anxiety around how responses might be judged. This may explain why confidence did not differ significantly from RV3, despite RV2's strengths. Thus, although RV2 was most preferred overall, it may also encourage self-censorship or over-editing in evaluative context.

\subsubsection{Both to Choose From (RV3): Freedom with Ambiguity}\leavevmode\\
The interface that allowed both editing and re-recording (RV3) received mixed feedback. Some participants appreciated its flexibility and valued having the choice to select the method that best suited their needs. This flexibility may have contributed to RV3’s highest confidence rating, significantly outperforming RV1. However, others reported feeling uncertain about which option to use and often second-guessed their decisions. This added an extra layer of friction, particularly in situations where the response felt especially important. Although participants could only use one option per attempt, some described thinking through both before making a decision, which added to their cognitive load. The absence of system guidance or prompts left users to figure out their own approach during the task. In this way, RV3’s flexibility came with increased mental effort which is supported by RV3’s significantly lower self-representation score compared to RV2, despite offering both options. \oc{As P78 put in, \textit{``The flexibility was reassuring, even if I had to think a bit more''}.}

\subsection{\oc{Response Interface Variants: Summary of Quantitative and Qualitative Analysis}}\leavevmode
\oc{Overall, we found that RV2 scored highest on control and expressiveness, and participants attributed this to the ease of fixing small mistakes without repeating entire responses, which made them feel both efficient and well represented. Its confidence score did not exceed RV3 because some users over edited out of concern for how they might appear, adding additional pressure on themselves despite the tool’s strengths. RV1 showed the lowest confidence and control, which matches participants’ self reports that mandatory re-recording created fatigue and led them to accept imperfect responses simply to avoid extra effort. RV3 achieved the highest confidence because participants liked having a choice, yet its lower expressiveness reflects the uncertainty users described when deciding between editing and re-recording.}

\subsection{Feedback Interface Variants: Quantitative Analysis}
We evaluated the effect of three feedback variants: FV1 (Motivational), FV2 (Evaluative), and FV3 (Both) on five outcome variables: perceived usefulness, confidence, engagement and, acknowledgment. Similar to RV analysis, Welch ANOVA was used to test for differences across groups. The results are presented in Table~\ref{tab:feedback_effects}.

We found statistically significant differences for usefulness ($F(2, 55) = 3.54$, $p = .036$, $\eta^2 = .107$), confidence ($F(2, 51) = 3.79$, $p = .029$, $\eta^2 = .064$), and acknowledgment ($F(2, 57) = 6.30$, $p = .003$, $\eta^2 = .138$). According to the effect sizes, these differences were moderate to large, indicating that the type of feedback meaningfully influenced user perceptions. No significant differences were found for engagement ($p = .241$), suggesting this aspect was less sensitive to variation in feedback design variations.

Post hoc Games–Howell comparisons revealed that for usefulness, participants rated FV3 (Both) significantly higher than FV2 (Evaluative) ($p = .031$), while FV1 (Motivational) also showed a marginal advantage over FV2 ($p = .056$). For confidence, FV1 was marginally preferred over FV3 ($p = .055$), indicating that motivational feedback alone may be perceived as more effective than when combined with evaluative cues to enhance competence. For acknowledgment, FV3 was rated significantly higher than FV2 ($p = .002$), highlighting the benefit of combining motivational and evaluative elements for offering a sense of being valued.

No significant pairwise differences were found for engagement, and the mean values were relatively similar across all feedback variants. This suggests that the type of feedback provided may not have a strong influence on enhancing participants’ sense of engagement to perform better.

\begin{table*}[ht]
\centering
\footnotesize
\renewcommand{\arraystretch}{1.3}
\setlength{\tabcolsep}{5pt}
\begin{tabular}{|l|c|c|c|p{3.6cm}|}
\hline
\textbf{Variable} & \textbf{FV1 (Motivational)} & \textbf{FV2 (Evaluative)} & \textbf{FV3 (Both)} & \textbf{Post-Hoc Analysis} \\
\hline
\makecell[l]{\textbf{Usefulness} \\ $F(2,55) = 3.54$, \textbf{$p = .036$}} & 
M = 3.467, SD = 0.507 &
M = 2.967, SD = 1.033 &
\textbf{M = 3.533}, SD = 0.571 &
FV3 > FV2* \\
\hline
\makecell[l]{\textbf{Confidence} \\ $F(2,51) = 3.79$, \textbf{$p = .029$}} & 
\textbf{M = 3.300}, SD = 0.535 &
M = 2.933, SD = 0.980 &
M = 2.667, SD = 1.348 &
FV1 > FV3 (marginal significance; $p < .06$) \\
\hline
\makecell[l]{\textbf{Engagement} \\ $F(2,58) = 1.46$, $p = .241$} & 
M = 3.333, SD = 0.661 &
M = 3.300, SD = 0.794 &
M = 3.033, SD = 0.765 &
— \\
\hline
\makecell[l]{\textbf{Acknowledgment} \\ $F(2,57) = 6.30$, \textbf{$p = .003$}} & 
M = 3.100, SD = 0.662 &
M = 2.700, SD = 0.837 &
\textbf{M = 3.400}, SD = 0.675 &
FV3 > FV2** \\
\hline
\end{tabular}
\caption{Welch ANOVA and Games–Howell post hoc comparisons across feedback variants. Significant post-hoc differences are marked with asterisks (* $p < .05$, ** $p < .01$, *** $p < .001$).}
\label{tab:feedback_effects}
\end{table*}

\subsection{Feedback Interface Variants: Qualitative Analysis}
\subsubsection{Motivational Feedback (FV1): Emotional Safety Interpreted as Presence}\leavevmode\\
User responses to the motivational feedback condition indicated a consistent perception of emotional support and attentiveness. Although this variant offered only affirmation and lacked performance-related critique, participants rated it highly on perceived confidence. This pattern suggests that motivational feedback was not interpreted as superficial or generic. Rather, the absence of criticism was perceived as intentional and supportive. Beyond affective reception, it may also have encouraged greater self-regulation. By not prescribing specific changes, the system implicitly placed the responsibility on the user to reflect and adjust on their own terms. In this way, the design may have supported autonomy by offering space for self-directed improvement rather than externally guided correction. \oc{As P107 described \textit{``It (feedback) felt like someone cheering me on while I was answering [...]''} and P116 found it \textit{``warm and supportive''} reinforcing the sense of emotional alignment.}

\subsubsection{Evaluative Feedback (FV2): Constructive Yet Emotionally Distant}\leavevmode\\
This variant delivered targeted, human-like suggestions after each response and was clearly personalized at the task level. However, it was rated lowest in acknowledgment and usefulness. While the feedback was informative, its tone appeared to prioritize correction over connection. \oc{Participants (P124, P139 \& P147) described it as \textit{``critical''}, indicating that the system’s emphasis on identifying areas for improvement may have diminished the sense of relational presence. P141 mentioned \textit{``It gave useful suggestions but [...] felt more like an evaluation than a conversation''}.} Despite being constructive, the feedback may have been perceived as judgmental rather than supportive. The experience of repeated critique, even when phrased with care, appeared to create a sense of being monitored rather than guided. In offering direction, the design may have compromised users’ sense of recognition. 

\subsubsection{Both Feedback Types (FV3): Acknowledging Without Anchoring}\leavevmode\\
The combined motivational and evaluative feedback after each response received the highest ratings in acknowledgment and usefulness. Although it suggests that users felt both recognized and guided, its perceived confidence score was lower than motivational feedback, indicating some ambiguity in how the feedback was interpreted. While the balance of encouragement and critique may have helped reduce emotional strain, it may also have made the system’s intent less clear. Notably, this condition received the lowest mean score in engagement. \oc{Although the difference was not statistically significant, the trend suggests that incorporating evaluative elements may have diluted the emotional impact that motivational feedback alone achieved, as reported by P159: \textit{``supportive, but also pointed out too many things''}. At the same time, the strong ratings in acknowledgment and usefulness indicate that combining critique with support may have addressed a broader range of user expectations as P179 commented, \textit{``It gave me direction without feeling harsh''}. Also, participants (P162 \& P171) perceived it as a form of \textit{``gentle criticism''} and beneficial for their self-improvement.} 

\subsection{\oc{Feedback Interface Variants: Summary of Quantitative and Qualitative Analysis}}
\oc{We found that FV3 scored highest on usefulness and acknowledgment, and participants attributed this to receiving both encouragement and concrete guidance, which made them feel recognized and supported while still knowing how to improve. Its confidence score did not exceed FV1 because some users felt that mixing critique with praise created uncertainty about the system’s intent, adding mild interpretive pressure despite its strong benefits. FV2 showed the lowest usefulness and acknowledgment, which aligns with participants comments that predominantly evaluative messages felt critical and emotionally distant, leading them to feel corrected rather than valued. FV1 achieved the highest confidence because participants interpreted motivational feedback as warm and attentive, yet its lower usefulness reflects the limited direction users described when the system offered affirmation without specific suggestions.}


\section{Discussion}

\subsection{Study 1: Performing for the Machine at the Cost of Authenticity}
Participants approached AI interviews with expectations shaped by exposure to current AI-mediated technologies and by inflated organizational terminology. Familiarity with tools like ChatGPT facilitated expectations of responsiveness and intelligence, while corporate terms such as ``AI recruiter'' implied advanced, conversational systems. However, the reality differed: static interfaces, no feedback, and rigid processes that felt indifferent rather than adaptive. This gap reflects Expectation Violation Theory (EVT), in which anticipated social dynamics fail to occur, producing frustration, detachment, or mistrust \cite{burgoon2015expectancy, grimes2021mental}. Such misalignment often leads to negative reactions, especially when systems fail to meet expectations set by their design or framing \cite{langer2022look, roemmich2023values}.

\oc{The disappointment was not only about functionality but also about recognition. Participants carefully managed their impressions to match imagined algorithmic preferences. Studies show that algorithmic assessments frequently limit self-expression and ignore individual context \cite{aizenberg2025examining}. Many began viewing the process as theatrical, even dehumanizing; ``not even real,'' one Redditor said. Others questioned whether the interview was a legitimate evaluation or merely a pipeline for training data. This skepticism echoes broader findings that candidates often feel emotionally misjudged or unseen by automated systems, especially in asynchronous formats \cite{pyle2024us, langer2020anybody}. The lack of transparency and organizational silence, including ghosting, deepened this suspicion. Ghosting is not uncommon in hiring, but in this context, where no human presence was ever felt, it did not just signal rejection; it made participants feel invisible, as if their effort had never mattered at all. Such dehumanizing perceptions are well documented in responses to emotion AI and hiring automation, where the absence of human acknowledgment leads to moral disengagement and reduced trust \cite{roemmich2023values, park2021human}. In response, applicants described using AI tools to script answers, altering speech delivery, or uploading polished audio. This behavior reflects strategic adaptation to the interviewer algorithm, aligning with prior research on workarounds and appropriation in sociotechnical systems \cite{doggett2024digital, davison2019adopted}.}

For neurodivergent participants, the stakes were higher. Research shows that marginalized users often hide identity traits to avoid misinterpretation by algorithmic systems, which contributes to systemic inequity \cite{van2022intersectional, suen2025comparing}. \oc{If applicants feel compelled to mask their neurodivergence or find workarounds to appear ``neutral,'' the system no longer supports equity; it instead rewards conformity. The case of AI interviews adds another dimension: systems designed for efficiency and standardization can inadvertently enforce a ``one-size-fits-all'' interaction style, which can be challenging for neurodivergent users and others whose communication practices diverge from the norm.} Prior work underscores the need to design AI systems that are inclusive of neurodivergent users, not just accessible in theory but also responsive in practice \cite{van2022intersectional}.

\subsection{Study 2: Designing System Toward Equitable Experiences}

\oc{Our findings highlight how user agency and transparency shape candidate experience in AI-mediated interviews, particularly through response control and feedback. Response variants provide candidates with some control over self-presentation, but this control is uneven and often difficult to interpret. Re-recording allows candidates to start over; however, it can feel demanding, as candidates must improve responses without interactional guidance. Prior work shows that many challenges in asynchronous AI interviews stem from the absence of interactional cues that signal understanding and evaluative stance in human interviews \cite{rivera2015go}. Without in-the-moment feedback, candidates lack information about how their responses are received, which limits their ability to adjust or clarify and reduces perceived control.}

\oc{Editing within a limited window offers a more bounded form of agency, allowing candidates to refine responses without discarding them entirely. However, flexibility alone does not ensure a sense of control. When options are not clearly framed, flexibility can increase uncertainty \cite{adnin2022hiring, aizenberg2025examining}, as candidates question which choice aligns with system expectations. Prior work suggests that candidates often seek clearer opportunities to explain intent or provide context, such as clarifying circumstances or technical issues, rather than unlimited control \cite{pyle2024us}. This points to design directions that emphasize transparent and constrained forms of agency, where candidates understand what can be changed and how those changes are used.}

\oc{Feedback further shapes perceptions of control and openness. Motivational feedback was often experienced as supportive, while evaluative feedback provided direction but could feel intrusive if delivered during interaction. Perhaps, feedback is better received after the interview, where it supports reflection rather than immediate judgment \cite{panadero2022review, sanches2019hci, mieczkowski2021ai}. From a transparency perspective, feedback signals how the system and organization interpret candidate responses. Generic feedback may offer reassurance, but more contextual feedback can help candidates better understand evaluation standards and outcomes \cite{ananny2018seeing}.}

\oc{Recruiters and HR stakeholders further emphasize that organizational practices play a significant role in shaping how agency and transparency are experienced in AI mediated interviews. While these systems are often valued for efficiency, stakeholders acknowledge that candidates have limited visibility into how responses are assessed, particularly in asynchronous settings \cite{lashkari2023finding, torres2017interviews}. As a result, candidates often rely on inference rather than explicit communication to understand expectations or next steps in the process \cite{ilhan2025opportunities}. Stakeholders note that decisions about response flexibility and feedback are typically determined through employer controlled system settings, such as whether revisions are allowed and how many attempts are permitted. Although these configurations are intended to balance consistency and fairness, the reasoning behind them and their role in evaluation are rarely communicated to candidates, which can make it difficult for candidates to understand the boundaries of their control.}

\oc{At the same time, employers commonly frame AI interview systems as tools that support, rather than replace, human decision making \cite{raghavan2020mitigating}. From the candidate perspective, however, this distinction is not always apparent during the interview itself. Prior organizational research suggests that how AI systems are embedded in practice influences how judgments are formed and how accountability is perceived, indicating that transparency depends not only on system design but also on how organizations communicate their use of these tools \cite{parasuraman2010complacency, hirsch2017designing}. Taken together, these perspectives suggest that user agency and transparency in AI mediated interviews are shaped through both interface features and organizational communication. Stakeholders therefore point to the need for approaches that make the boundaries of system control more visible and that more clearly communicate expectations, interpretation, and follow up, rather than relying on technical improvements alone \cite{latif2022ai, felz2022ai}.}

\subsection{Overall Discussion: From Inflated Expectations to Algorithmic Sensemaking}

\oc{The growing advancement of technologies has raised the bar of user expectations. As interactions with advanced systems become routine, users internalize certain standards, making it difficult to accept systems that fall short. With increasing over-reliance on and extensive use of LLM-driven systems \cite{jakesch2023co, he2024ai, kapania2025m}, users are reshaped to expect highly conversational and personalized interactions. This marks a shift in anthropomorphism, where users increasingly compare automated systems not to humans but to advanced language models. As others have observed, users may place undue trust in LLMs when they appear more human, influencing perceptions of credibility and authority \cite{cohn2024believing}. Expectation and interpretation often arise from one-sided experiences, where users approach systems with preformed assumptions that become more pronounced in the absence of transparent design. The interpretation of black-box systems is often shaped by subjective experience when objective explanations are limited or unavailable \cite{guidotti2018survey, armstrong2023navigating, fong2017interpretable}. While such subjectivity may appear trivial, our findings show that it can lead to serious trust breakdowns, such as when system opacity and rigidity led applicants to suspect the platform was a data-harvesting scam rather than a legitimate interview system.}

\oc{Beyond conversational use, AI-driven systems support a wide range of functions across domains. When organizations report using AI-driven systems, users often associate them primarily with interaction, even though such systems may be deployed solely for analysis or decision support \cite{roy2024exploring, wang2025lata}. Adopting AI-based technologies also remains costly for many small and mid-sized enterprises \cite{deloitte_cost}. While these constraints are understandable, organizations are still expected to demonstrate care toward applicants, which is often absent in practice despite promises to the contrary. Prior research shows that hiring services frequently make unsupported claims about solving hiring-related problems through their technologies, many of which do not hold up under scrutiny \cite{roemmich2023values}, and that such tools are often deployed without thorough evaluation \cite{shankar2024we}. In this context, the findings illustrate how candidates adapt to uncertainty by forming folk theories about what the system values, adjusting tone and wording to perform for an unseen algorithm, consistent with prior work on algorithmic sensemaking \cite{devito2021adaptive, xu2024folk}. Over time, this adaptation can narrow acceptable forms of self-presentation, encouraging safe and generic responses while discouraging personal or creative expression, echoing concerns that opaque systems promote risk-averse behavior \cite{binns2018s}. These dynamics heighten the risk of distrust and unmet expectations driven by inflated claims \cite{adnin2024look}, and in some cases may also perpetuate bias.}

\oc{The findings further point to the limits of interface-level interventions in shaping candidate experience. While edit options reduced cognitive and emotional burden by lowering the perceived risk of being candid, performance-oriented strategies largely persisted, suggesting that interaction design alone may not fully address deeper concerns about evaluation and recognition. Prior ACM research shows that perceptions of fairness and legitimacy in algorithmic decision-making depend less on transparency alone and more on signals of oversight, contestability, and accountability \cite{lee2015working}. At the same time, organizational constraints and efficiency pressures limit how much agency AI-driven hiring systems can realistically offer applicants. Given that applicants occupy a less powerful position and often operate under constrained conditions, responsibility falls more heavily on organizations to ensure that hiring processes enable fair and authentic performance \cite{lavanchy2023applicants}. More broadly, AI-driven interview tools are expected to support fairness, consistency, and accessibility. CSCW research has long emphasized empowering users to recognize and resist manipulative patterns and designing systems that uphold justice by treating individuals fairly across identities and contexts \cite{min2024avatar, acikgoz2020justice, pyle2024us, sun2023care}. Embedding these values into system design is therefore not merely a theoretical concern, but a practical requirement for building sustainable and equitable hiring technologies.}

\section{Limitations and Future Work}
While our work offers important contributions to the study of AI-driven hiring, it is not without limitations. First, although we analyzed a large volume of \textit{Reddit} data, the authenticity of these posts could not be independently verified. Similarly, our user studies relied on self-reported data, which may be subject to bias or inaccuracies. Second, the majority of our participants were students, which limits the generalizability of our findings to broader populations. Third, Study 2 involved a simulated interview scenario, which may not fully reflect real-world stakes or pressures. The transparency provided during the study may also have influenced participants’ responses in ways that differ from commercial deployments. Fourth, our work centers primarily on applicant perspectives, omitting the organizational viewpoint. Understanding how employers implement, interpret, and evaluate these systems remains an important area for future work. Fifth, while we identified several tools marketed as AI-driven based on company websites, we could not verify whether the organizations using them employed the same versions or configurations.

In future work, we plan to recruit professionals with industry experience, hiring managers and organizations to deepen our understanding of AI interview practices in operational contexts. We also intend to include participants from more diverse age groups and occupations to increase ecological validity. 

\section{Conclusion}
This paper advances the understanding of AI-driven asynchronous interviews by examining how job seekers experience and adapt to these systems amid the growing public familiarity with LLMs. Through large-scale analysis of \textit{Reddit} discussions and follow-up interviews, we uncovered critical mismatches between system design and user expectations mismatches often rooted in organizational rhetoric, anthropomorphic interface cues, and evolving assumptions about AI capabilities. These mismatches manifested as diminished trust, reduced perceived autonomy, and a sense of disconnection, among both neurodivergent and neurotypical users. Building on these findings, we designed an interview interface and evaluated based on Self-Determination Theory (SDT). By offering different configurations of response control and feedback, we explored how design can influence users’ sense of agency, competence, and acknowledgment. Our results demonstrate that simple yet intentional features such as editing options and motivational feedback can significantly shape candidate experience, while overly complex or ambiguous combinations may undermine user confidence. Together, these findings underscore the importance of expectation management, sense of agency, and psychological support in the design of AI hiring systems.

\bibliographystyle{ACM-Reference-Format}
\bibliography{sample-base}

\appendix

\section{Keywords used to filter Reddit comments}
\label{keywords}

AI interviewer \cite{liu2023speech}, AI interviews \cite{mcgee2023ai}, AI-driven interview \cite{sumathi2024ai}, automated interview \cite{langer2020highly}, AI-based video interview \cite{suen2023building}, Algorithmic hiring \cite{raghavan2020mitigating}, AI tools in recruitment \cite{horodyski2023recruiter}, AI-based interview \cite{lee2021development}, AI-powered interviews \cite{kothari2024interviewease, biswas2024hi}, AI-driven hiring \cite{mangal2023analytical}, automation in recruitment \cite{gupta2018automation}, virtual interview \cite{jones2020virtual}, asynchronous video interview \cite{lukacik2022into}, video based interview \cite{alonzo2016declarative}, one way video interview \cite{poh2015evaluating}

\section{Interview Questionnaire}
\label{int_qs}
\begin{itemize}
    \item Can you share your overall experience with AI interviewers?
    \item Did the AI interview tool meet your expectations? If not, what was missing or different from what you had anticipated?
    \item Did you receive any guidance or instructions from the organization on how to prepare for the AI interview?
    \item How confident did you feel while responding to the AI interviewer? Were there any particular reasons for feeling confident or unsure?
    \item Did you face any technical difficulties, such as internet issues, lag, or audio/video problems? How did you manage or resolve them?
    \item Were there any issues with how long you had to respond or how your responses were timed?
    \item Were you able to review or re-record your answers, or did you feel pressured to get it right in one take? Would the ability to edit your responses have changed your experience?
    \item What is your impression of how AI systems evaluate candidates? Do you have any thoughts or assumptions about how the decisions are made?
    \item Did you have any concerns about potential bias in the system,for example, based on your accent, appearance, neurotype, or background?
    \item Before the interview, were you aware of which AI tool or platform you’d be using?
    \item Did the interview questions feel personalized or adaptive to you, or did they seem generic and one-size-fits-all?
    \item Some people say the questions in AI interviews are short and lacking in depth. Did you experience something similar?
    \item While recording your responses, did you encounter any challenges (e.g., time pressure, speaking naturally, staying focused)? How did you deal with them?
    \item Did the AI interviewer simulate human-like behavior in any way (e.g., facial cues, verbal responses, timing)? If so, how did that affect your experience?
    \item Were you expected to maintain specific facial expressions, look into the camera, or behave in a certain way during the recording?
    \item From your perspective, what was one important element missing from the AI interview process?
    \item How transparent did the AI interview feel to you? Were you informed about how your responses would be processed or evaluated?
    \item Was the interview recorded, and were you notified whether your facial expressions, voice, or other biometric data would be analyzed?
    \item Did the overall process feel respectful and fair, or did it feel impersonal or dismissive in any way?
    \item Were there any parts of the interview that felt inaccessible or difficult to navigate due to your personal or neurodivergent needs?
    \item Overall, were you satisfied with the AI interview experience? What would you change or improve?
    \item Compared to a human interviewer, did you feel more or less comfortable facing the AI system? Why?

\end{itemize}

\section{Few-Shot Prompting to Identify Relevant Reddit Data}
\label{prompting}

System Prompt: You are a helpful assistant trained to identify whether a Reddit data point describes a person’s experience with AI-based asynchronous interviews. These interviews are typically one-way, where applicants face AI-based interview without a human interviewer.\\

User Prompt: Below are several Reddit data points. For each entry, respond with either "Relevant" or "Irrelevant", followed by a short justification explaining your decision. A relevant data point should mention or strongly imply personal experience (positive, negative or neutral) with asynchronous interviews that may or may not involve AI tools, such as HireVue or similar systems. Followings are a few examples:\\

Examples:\\
Comment 1:\\
``I had to stare at my webcam and talk to a bot for 20 minutes. It was exhausting and felt like talking into the void.''\\
Response:\\
Relevant – This comment clearly refers to a one-way AI-based interview experience.\\

Comment 2:\\
``My friend told me these AI interviews are just a scam. I haven’t done one myself.''\\
Response:\\
Irrelevant – The speaker did not participate in the interview themselves.\\

Comment 3:\\
"I’ve used Zoom for a few interviews. Nothing fancy."\\
Response:\\
Irrelevant – This refers to synchronous, human-led interviews.\\

\section{Behavioral Questions}
\label{behave_qs}
These following questions are collected from \textit{Indeed} \cite{indeed_qs}. \\

\textbf{Used in Response Variants:}\\
1. Give me an example of a time you had a conflict with a team member. How did you handle it?\\
2. Tell me about a time you made a mistake at work. How did you resolve the problem, and what did you learn from your mistake? \\
3. Describe an occasion when you had to use time management skills to complete a task or meet a deadline. How did you do it? \\

\textbf{Used in Feedback Variants:}\\
1. Describe an occasion when you failed at a task. What did you learn from it?\\
2. Tell me about a time you took the initiative in your career. What was your motivation for doing so?\\
3. Describe a time when you used your leadership skills to motivate your team or colleagues.

\section{Few-Shot Prompting To Generate Feedback}
\label{prompt_feedback}
\subsection{Motivational Feedback}
You are a supportive mentor giving motivational feedback to someone who is responding to behavioral interview responses. Your job is to:

\begin{itemize}
    \item Encourage their effort
    \item Reinforce their self-reflection
    \item Avoid any kind of judgment, critique, or correction
    \item Be warm, affirming, and growth-oriented
    \item Keep the tone friendly and inspiring
    \item Write in 1–3 natural, uplifting sentences
\end{itemize}

Use the examples below as guidance:

\textbf{Example 1}

\textbf{Q1:} Give me an example of a time you had a conflict with a team member. How did you handle it?\\
\textbf{Response:} ``I had a disagreement with a teammate about how to divide tasks. We had different views on priorities, but eventually we talked it through and came to a compromise.''

\textbf{Motivational Feedback:} \textit{``Thank you for sharing this experience. It takes courage to reflect on conflict. You're clearly engaging with the story and showing awareness, and that's an important step forward. Keep practicing and trusting your instincts, you’re doing great.''}

\textbf{Example 2}

\textbf{Q2:} Tell me about a time you made a mistake at work. How did you resolve the problem, and what did you learn from your mistake?\\
\textbf{Response:} ``I forgot to send out an important email to a client. Once I realized, I contacted them right away to apologize and sent the information. It made me more careful with deadlines in the future.''

\textbf{Motivational Feedback:} \textit{``That’s a brave reflection, thanks for being open about it. It’s clear you care about doing well and learning from experiences. Keep it up, this kind of insight shows real growth.''}

\textbf{Example 3}

\textbf{Q3:} Describe an occasion when you had to use time management skills to complete a task or meet a deadline. How did you do it?\\
\textbf{Response:} ``I had to juggle multiple deadlines during finals week. I made a daily schedule, blocked out time for each task, and avoided distractions. I managed to submit everything on time.''

\textbf{Motivational Feedback:} \textit{``Well done ! You’re showing commitment and intention, and that’s something to be proud of. Taking time to reflect on how you managed it is a sign of strong self-awareness. Keep up the great work!''}

\subsection{Evaluative Feedback}

You are a thoughtful interviewer giving evaluative feedback on behavioral responses. For each response you read, write feedback that is:

\begin{itemize}
    \item \textbf{Friendly and natural in tone}
    \item \textbf{Evaluative and helpful}
    \item \textbf{Framed around the STAR method (Situation, Task, Action, Result)}
    \item \textbf{Suggestive, not judgmental, ike a helpful mentor or coach}
    \item \textbf{Written in 2–4 sentences}
\end{itemize}

Use these examples as a guide:\\

\textbf{Example 1}

\textbf{Q1:} Describe an occasion when you failed at a task. What did you learn from it?\\
\textbf{Response:} ``I was leading a small marketing campaign and forgot to confirm our ad schedule with the publishing team. We missed our planned launch window. I took responsibility and personally contacted the team to find an alternate slot, which we secured the following week. Since then, I always double-check timelines and confirmations.''

\textbf{Feedback:} \textit{``Thanks for sharing that candidly. You explained the situation and your role well, and your ownership is commendable. To strengthen this further, you could expand on what concrete strategies you now use to avoid similar mistakes,this helps make your learning more visible.''}

\textbf{Example 2}

\textbf{Q2:} Tell me about a time you took the initiative in your career. What was your motivation for doing so?\\
\textbf{Response:} ``During a slow quarter, I suggested we repurpose existing content into short-form videos for social media. I drafted scripts and coordinated with the design team to get them out quickly. Engagement went up by 30\%. I wasn’t told to do it, but I knew we needed to boost visibility.''

\textbf{Feedback:} \textit{``This is a solid example of taking initiative with a measurable impact. To deepen your answer, it might help to explain how you identified the opportunity in the first place and what feedback you received afterward. That kind of reflection highlights your awareness and influence.''}

\textbf{Example 3}

\textbf{Q3:} Describe a time when you used your leadership skills to motivate your team or colleagues.\\
\textbf{Response:} ``We were behind on a group project. I called a team huddle, broke down tasks into manageable parts, and created a progress board everyone could update. People became more engaged and we completed it on time.''

\textbf{Feedback:} \textit{``This shows thoughtful coordination and problem-solving under pressure. You might enhance it by sharing how your leadership approach impacted the team dynamic or whether it led to lasting change. Specific feedback from team members could also enrich the story.''}

\section{Tool Names Found in Our Study}
\label{tool_name}
HireVue, Sapia AI, Talview, Vervoe, XOR, Willo, SparkHire, Clovers AI, Skillora AI, Paradox AI, Eightfold AI, Echotalent AI, Vhire, MyInterview, Quinncia

\end{document}